\documentclass[amssymb,amsmath,prd,twocolumn,superscriptaddress,nofootinbib]{revtex4-1}

\usepackage{graphicx}
\usepackage{bm}
\usepackage{amssymb,amsmath}
\usepackage{mathrsfs}
\usepackage{latexsym}
\usepackage{color}
\usepackage[normalem]{ulem}
\usepackage{dcolumn}
\usepackage[colorlinks=true,citecolor=blue,urlcolor=blue]{hyperref}
\usepackage[usenames,dvipsnames]{xcolor}
\usepackage{xspace}
\usepackage{graphicx}
\usepackage{multirow}

\allowdisplaybreaks

\newcommand{\CIT}{\affiliation{Department of Physics, California Institute of Technology, Pasadena, California 91125, USA}}
\newcommand{\CITLab}{\affiliation{LIGO Laboratory, California Institute of Technology, Pasadena, California 91125, USA}}

\definecolor{kcmagenta}{rgb}{0.54, 0.17, 0.88}
\definecolor{shyellow}{rgb}{0.15625, 0.609375, 0.316406}
\definecolor{chorange}{rgb}{0.851, 0.372, 0.007}
\definecolor{tlteal}{rgb}{0,.55,.55}
\definecolor{jcpink}{rgb}{1.0, 0.0, 0.5}
\definecolor{mmgreen}{rgb}{0.0, 0.8, 0.6}
\definecolor{bbsalmon}{rgb}{1.0, 0.47, 0.42}

\newcommand{\chieff}{\chi_{\textrm{eff}}}
\newcommand{\chip}{\chi_{p}}
\newcommand{\flow}{f_{\textrm{low}}}
\newcommand{\Mc}{\mathcal{M}}

\begin{document}

\title{The curious case of GW200129: interplay between spin-precession inference and data-quality issues}

\author{Ethan Payne} \CIT \CITLab
\author{Sophie Hourihane} \CIT \CITLab
\author{Jacob Golomb} \CIT \CITLab
\author{Rhiannon Udall} \CIT \CITLab
\author{Derek Davis} \CIT \CITLab
\author{Katerina Chatziioannou} \CIT \CITLab 

\date{\today}

\begin{abstract}
Measurement of spin-precession in black hole binary mergers observed with gravitational waves is an exciting milestone as it relates to both general relativistic dynamics and astrophysical binary formation scenarios. In this study, we revisit the evidence for spin-precession in GW200129 and localize its origin to data in LIGO Livingston in the 20--50\,Hz frequency range where the signal amplitude is lower than expected from a non-precessing binary given all the other data. These data are subject to known data quality issues as a glitch was subtracted from the detector's strain data. The lack of evidence for spin-precession in LIGO Hanford leads to a noticeable inconsistency between the inferred binary mass ratio and precessing spin in the two LIGO detectors, something not expected from solely different Gaussian noise realizations. 
We revisit the LIGO Livingston glitch mitigation and show that the difference between a spin-precessing and a non-precessing interpretation for GW200129 is smaller than the statistical and systematic uncertainty of the glitch subtraction, finding that the support for spin-precession depends sensitively on the glitch modeling. We also investigate the signal-to-noise ratio $\sim7$ trigger in the less sensitive Virgo detector. Though not influencing the spin-precession studies, the Virgo trigger is grossly inconsistent with the ones in LIGO Hanford and LIGO Livingston as it points to a much heavier system. We interpret the Virgo data in the context of further data quality issues. While our results do not disprove the presence of spin-precession in GW200129, we argue that any such inference is contingent upon the statistical and systematic uncertainty of the glitch mitigation. Our study highlights the role of data quality investigations when inferring subtle effects such as spin-precession for short signals such as the ones produced by high-mass systems.
\end{abstract}

\maketitle

%%%%%%%%%%%%%%%%%%%%%%%%%%%%%%%%%%%%%%%%%%%
\section{Introduction}
\label{sec:intro}
%%%%%%%%%%%%%%%%%%%%%%%%%%%%%%%%%%%%%%%%%%%

GW200129\_065458 (henceforth GW200129) is a gravitational wave (GW) candidate reported in GWTC-3~\cite{gwtc3}. The signal was observed by all three LIGO-Virgo detectors~\cite{TheLIGOScientific:2014jea,TheVirgo:2014hva} operational during the third observing run (O3) and it is consistent with the coalescence of two black holes (BHs) with source-frame masses $34.5^{+9.9}_{-3.2}\,M_{\odot}$ and $28.9^{+3.4}_{-9.3}\,M_{\odot}$ at the 90\% credible level. Though the masses are typical within the population of observed events~\cite{LIGOScientific:2021psn}, the event's signal-to-noise-ratio (SNR) of $26.8^{+0.2}_{-0.2}$ makes it the loudest binary BH (BBH) observed to date. Additionally, it is one of the loudest triggers in the Virgo detector with a detected SNR of 6--7 depending on the detection pipeline~\cite{gwtc3}. The signal temporally overlapped with a glitch in the LIGO Livingston detector, which was subtracted using information from auxiliary channels~\cite{Davis:2022ird}. The detection and glitch mitigation procedures for this event are recapped in App.~\ref{appendix:glitchsubtraction}.

The interpretation of some events in GWTC-3 was impacted by waveform systematics, with GW200129 being one of the most extreme examples. As part of the catalog, results were obtained with the {\tt IMRPhenomXPHM}~\cite{Pratten:2020ceb} and {\tt SEOBNRv4PHM}~\cite{Ossokine:2020kjp} waveform models using the parameter inference algorithms {\tt Bilby}~\cite{Ashton:2018jfp,Romero-Shaw:2020owr} and {\tt RIFT}~\cite{Wysocki:2019grj} respectively. Both waveforms correspond to quasicircular binary inspirals and include high-order radiation modes and the effect of relativistic spin-precession arising from interactions between the component spins and the orbital angular momentum. All analyses used the glitch-subtracted LIGO Livingston data. The {\tt IMRPhenomXPHM} result was characterized by large spins and a bimodal structure with peaks at $\sim0.45$ and $ \sim0.9$ for the binary mass ratio. The {\tt SEOBNRv4PHM} results, on the other hand, pointed to more moderate spins and near equal binary masses. Both waveforms, however, reported a mass-weighted spin aligned with the Newtonian orbital angular momentum of $\chieff \sim0.1$, and thus the inferred large spins with {\tt IMRPhenomXPHM} corresponded to spin components in the binary orbital plane and spin-precession. Such differences between the waveform models are not unexpected for high SNR signals~\cite{Purrer:2019jcp}. Waveform systematics are also likely more prominent when it comes to spin-precession, as modeling prescriptions vary and are not calibrated to numerical relativity simulations featuring spin-precession~\cite{Pratten:2020ceb, Pratten:2020igi, Ossokine:2020kjp}. Data quality issues could further lead to evidence for spin-precession~\cite{Ashton:2021tvz}. Due to differences in the inference algorithms and waveform systematics, GWTC-3 argued that definitive conclusions could not be drawn regarding the possibility of spin-precession in this event~\cite{gwtc3}. 

Stronger conclusions in favor of spin-precession~\cite{Hannam:2021pit} and a merger remnant that experienced a large recoil velocity~\cite{Varma:2022pld} were put forward by means of a third waveform model. {\tt NRSur7dq4}~\cite{Varma:2019csw} is a surrogate to numerical relativity simulations of merging BHs that is also restricted to quasicircular orbits and models the effect of high-order modes and spin-precession. This model exhibits the smallest mismatch against numerical relativity waveforms, sometimes comparable to the numerical error in the simulations. It is thus expected to generally yield the smallest errors due to waveform systematics~\cite{Varma:2019csw}. This fact was exploited in Hannam \textit{et al.}~\cite{Hannam:2021pit} to break the waveform systematics tie and argue that the source of GW200129 exhibited relativistic spin-precession with a primary component spin magnitude of $\chi_1=0.9^{+0.1}_{-0.5}$ at the 90\% credible level.  

During a binary inspiral, spin-precession is described through post-Newtonian theory~\cite{Apostolatos:1994mx,Kidder:1995zr}. Spin components that are not aligned with the orbital angular momentum give rise to spin-orbit and spin-spin interactions that cause the orbit to change direction in space as the binary inspirals, e.g.,~\cite{Buonanno:2002fy,Buonanno:2004yd,Schmidt:2010it,Schmidt:2012rh,Hannam:2013oca,Chatziioannou:2016ezg,Chatziioannou:2017tdw,Gerosa:2015tea,Kesden:2014sla,Ramos-Buades:2020noq}. The emitted GW signal is modulated in amplitude and phase, and morphologically resembles the beating between two spin-aligned waveforms~\cite{Fairhurst:2019vut} or a spin-aligned waveform that has been ``twisted-up"~\cite{Schmidt:2010it,Schmidt:2012rh}. As the binary reaches merger, numerical simulations suggest that the direction of peak emission continues precessing~\cite{OShaughnessy:2012iol}. Parameter estimation analyses using {\tt NRSur7dq4} find that spins and spin-precession can be measured from merger-dominated signals for certain spin configurations~\cite{Biscoveanu:2021nvg}, however the lack of analytic understanding of the phenomenon means that it is not clear how such a measurement is achieved.

The main motivation for this study is to revisit GW200129 and attempt to understand how spins and spin-precession can be measured from a heavy BBH with a merger-dominated observed signal. In Sec.~\ref{sec:precession} we use {\tt NRSur7dq4} to conclude that the evidence for spin-precession originates exclusively from the LIGO Livingston data in the 20--50\,Hz frequency range, where the inferred signal amplitude is lower than what a spin-aligned binary would imply given the rest of the data. This range coincides with the known data quality issues described in App.~\ref{appendix:glitchsubtraction} and first identified in GWTC-3~\cite{gwtc3}. LIGO Hanford is consistent with a spin-aligned signal, causing an inconsistency between the inferred mass ratio $q$ and precession parameter $\chip$ inferred from each LIGO detector separately.
By means of simulated signals, we argue that such $q-\chip$ inconsistency is unlikely to be caused solely by the different Gaussian noise realizations in each detector at the time of the signal, rather pointing to remaining data quality issues beyond the original glitch-subtraction~\cite{gwtc3}. We also re-analyze the LIGO Livingston data above $50$\,Hz, (while keeping the original frequency range of the LIGO Hanford data) and confirm that all evidence for spin-precession disappears.

In the process, we find that the Virgo trigger, though consistent with a spin-aligned BBH, is \emph{inconsistent} with the signal seen in the LIGO Hanford and LIGO Livingston detectors. Specifically, the Virgo data are pointing to a much heavier BBH that merges $\sim$20\,ms earlier than the one observed by the LIGO detectors. We discuss Virgo data quality considerations in Sec.~\ref{sec:glitchV} within the context of a potential glitch that affects the inferred binary parameters if unmitigated. As a consequence, we do not include Virgo data in the sections examining spin-precession unless otherwise stated. The Virgo-LIGO inconsistency can be resolved if we use {\tt BayesWave}~\cite{Cornish:2014kda,Littenberg:2014oda,Cornish:2020dwh} to simultaneously model a CBC signal and glitches with CBC waveform models and sine-Gaussian wavelets respectively~\cite{Chatziioannou:2021ezd,Hourihane:2022doe}. The Virgo data are now consistent with the presence of both a signal that is consistent with the one in the LIGO detectors and an overlaping glitch with SNR $\sim4.6$. 

In Sec.~\ref{sec:glitchL} we revisit the LIGO Livingston data quality issues and compare the original glitch-subtraction based on {\tt gwsubtract}~\cite{Davis:2018yrz,Davis:2022ird} that uses information from auxiliary channels and the glitch estimate from {\tt BayesWave} that uses only strain data. Though the CBC model used in {\tt BayesWave} does not include the effect of spin-precession, we show that differences between the reconstructed waveforms from a non-precessing and spin-precessing analysis for GW200129 are \emph{smaller} than the statistical uncertainty in the glitch inference. Such differences can therefore not be reliably resolved in the presence of the glitch and its subtraction procedure. The two glitch estimation methods give similar results within their statistical errors, however {\tt gwsubtract} yields typically a lower glitch amplitude. We conclude that any evidence for spin-precession from GW200129 is contingent upon the systematic and statistical uncertainties of the LIGO Livingston glitch subtraction. Given the low SNR of the LIGO Livingston glitch and the glitch modeling uncertainties, we can at present not conclude whether the source of GW200129 exhibited spin-precession or not. 

In Sec.~\ref{sec:conclusions} we summarize our arguments that remaining data quality issues in LIGO Livingston cast doubt on the evidence for spin-precession. Besides data quality studies (i.e., spectrograms, glitch modeling, auxiliary channels), our investigations are based on comparisons between different detectors as well as different frequency bands of the same detector. We propose that similar investigations in further events of interest with exceptional inferred properties could help alleviate potential contamination due to data quality issues.

%%%%%%%%%%%%%%%%%%%%%%%%%%%%%%%%%%%%%%%%%%%
\section{The origin of the evidence for spin-precession}
\label{sec:precession}
%%%%%%%%%%%%%%%%%%%%%%%%%%%%%%%%%%%%%%%%%%%
%
\begin{figure*}
    \centering
    \includegraphics[width=0.49\linewidth]{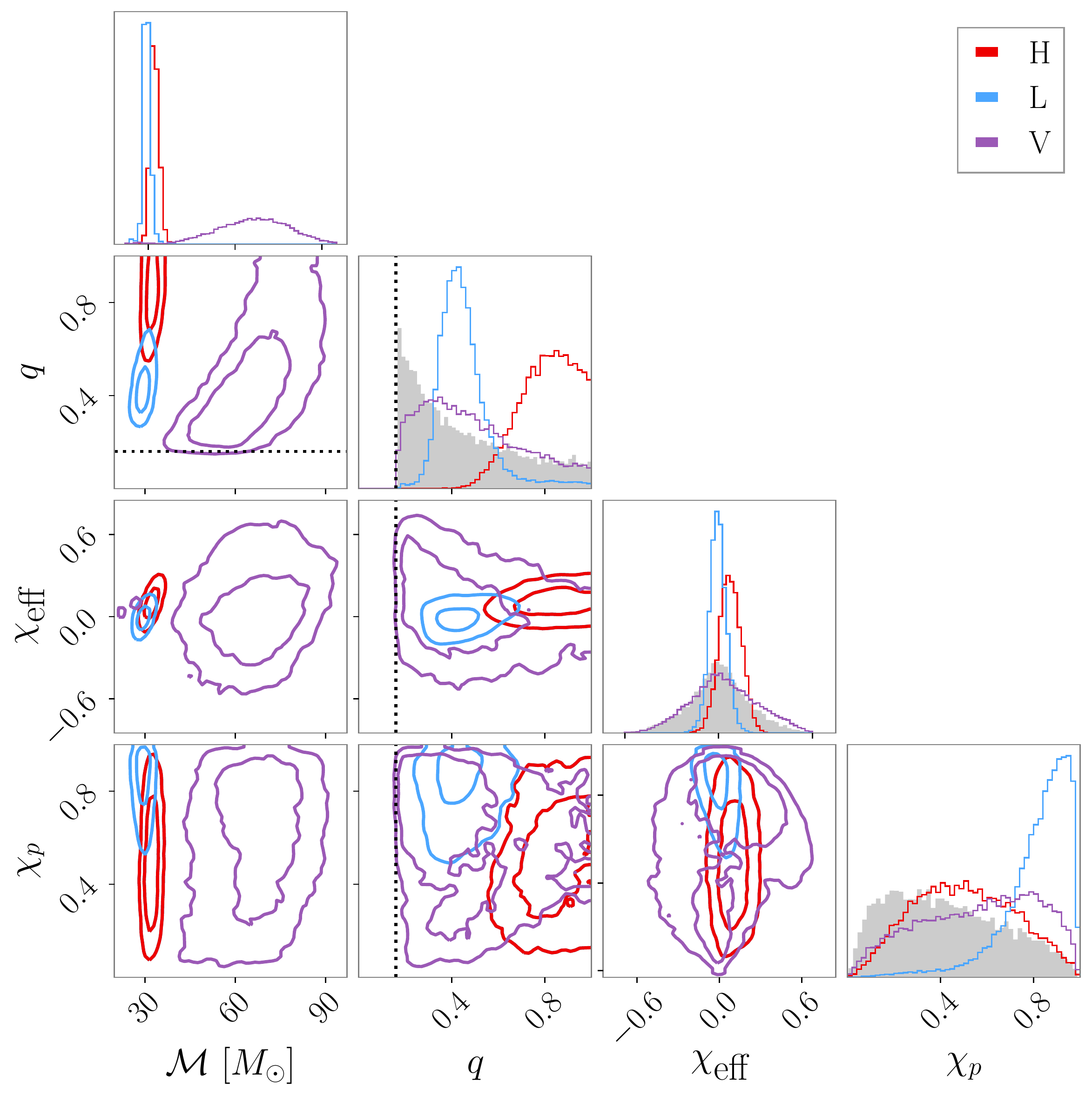}
    \includegraphics[width=0.49\linewidth]{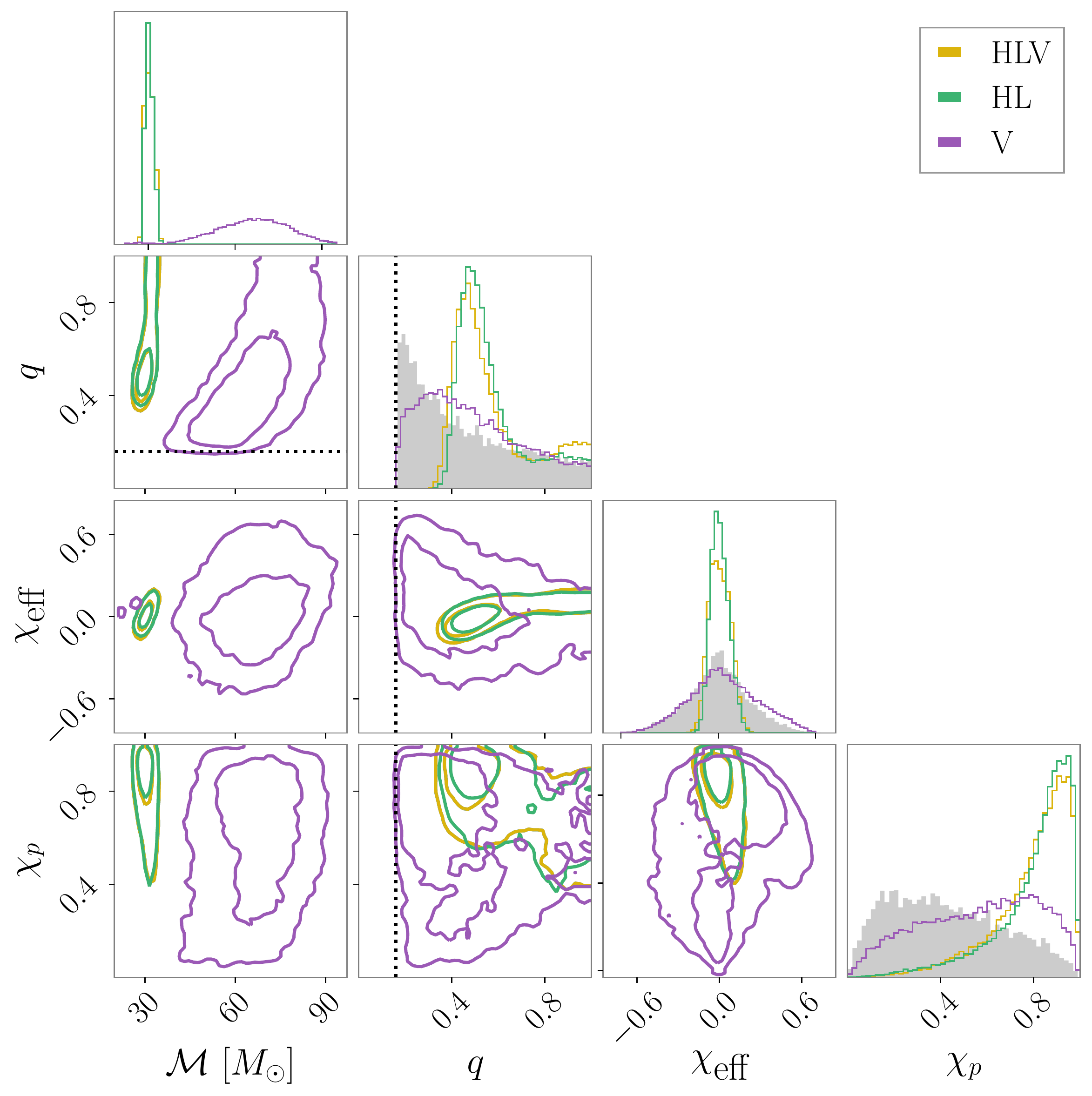}
    \caption{One- and two-dimensional marginalized posteriors for select intrinsic binary parameters: detector frame chirp-mass $\Mc$, mass ratio $q$, effective spin $\chieff$, and precessing spin $\chip$. See Table~\ref{tab:Bilbyrunsettings} for analysis settings and App.~\ref{appendix:bilby} for detailed parameter definitions. Two-dimensional panels show 50\% and 90\% contours. The black dashed line marks the minimum bound of $q$=1/6 in {\tt NRSur7dq4}'s region of validity. Shaded regions shows the prior for $q$, $\chieff$, $\chip$. The $\Mc$ prior increases monotonically to the maximum allowed value (see App.~\ref{appendix:bilby} for details on choices of priors). Left panel: comparison between analyses that use solely LIGO Hanford (red; H), LIGO Livingston (blue; L), and Virgo (purple; V) data. Right panel: comparison between analyses of all three detectors (yellow; HLV), only LIGO data (green; HL) and only Virgo data (purple; V). The evidence for spin-precession originates solely from the LIGO Livingston data as the other detectors give uninformative $\chip$ posteriors. Additionally, the binary masses inferred based on Virgo only are inconsistent with those from the LIGO data.}
    \label{fig:intrinsic}
\end{figure*}

Our main goal is to pinpoint the parts of the GW200129 data that are inconsistent with a non-precessing binary and understand the relevant signal morphology. Due to different orientations, sensitivities, and noise realizations, different detectors in the network do not observe an identical signal. The detector orientation, especially, affects the signal polarization content and thus the degree to which spin-precession might be measurable in each detector. Motivated by this, we begin by examining data using different detector combinations.

We perform parameter estimation using the {\tt NRSur7dq4} waveform and examine data from each detector separately (left panel) as well as the relation between the LIGO and the Virgo data (right panel) and show posteriors for select intrinsic parameters in Fig.~\ref{fig:intrinsic}. Analysis settings and details are provided in App.~\ref{appendix:bilby} and in all cases we use the same LIGO Livingston data as GWTC-3~\cite{gwtc3} where the glitch has been subtracted. Though we do not expect the posterior distributions for the various signal parameters inferred with different detector combinations to be identical, they should have broadly overlapping regions of support. If the triggers recorded by the different detectors are indeed consistent, any shift between the posteriors should be at the level of Gaussian noise fluctuations. 

The left panel shows that the evidence for spin-precession arises primarily from the LIGO Livingston data, whereas the precession parameter $\chip$ posterior is much closer to its prior when only LIGO Hanford or Virgo data are considered. A similar conclusion was reached in Hannam \textit{et al.}~\cite{Hannam:2021pit}. There is reasonable overlap between the two-dimensional distributions that involve the chirp mass $\mathcal{M}$, the mass ratio $q$, and the effective spin $\chieff$ inferred by the two LIGO detectors, as expected from detectors that observe the same signal under different Gaussian noise realizations. The discrepancy between the spin-precession inference in the two LIGO detectors, however, is evident in the $q-\chip$ panel. The two detectors lead to non overlapping distributions that point to either unequal masses and spin-precession (LIGO Livingston), or equal masses and no information for spin-precession (LIGO Hanford). 

Besides an uninformative posterior on $\chip$, the left panel points to a bigger issue with the Virgo data: inconsistent inferred masses. The right panel examines the role of Virgo in more detail in comparison to the LIGO data. Due to the lower SNR in Virgo, the intrinsic parameter posteriors are essentially identical between the HL and the HLV analyses. The lower total SNR means that the Virgo-only posteriors will be wider, but they are still expected to overlap with the ones inferred from the two LIGO detectors. However, this is not the case for the mass parameters as is most evident from the two dimensional panels involving the chirp mass. While the LIGO data are consistent with a typical binary with (detector-frame) chirp mass $30.3^{+2.5}_{-1.6}\,M_{\odot}$ at the 90\% credible level, the Virgo data point to a much heavier binary with $66.7^{+19.7}_{-22.6}\,M_{\odot}$ at the same credible level.

\begin{figure}
    \centering
    \includegraphics[width=0.49\textwidth]{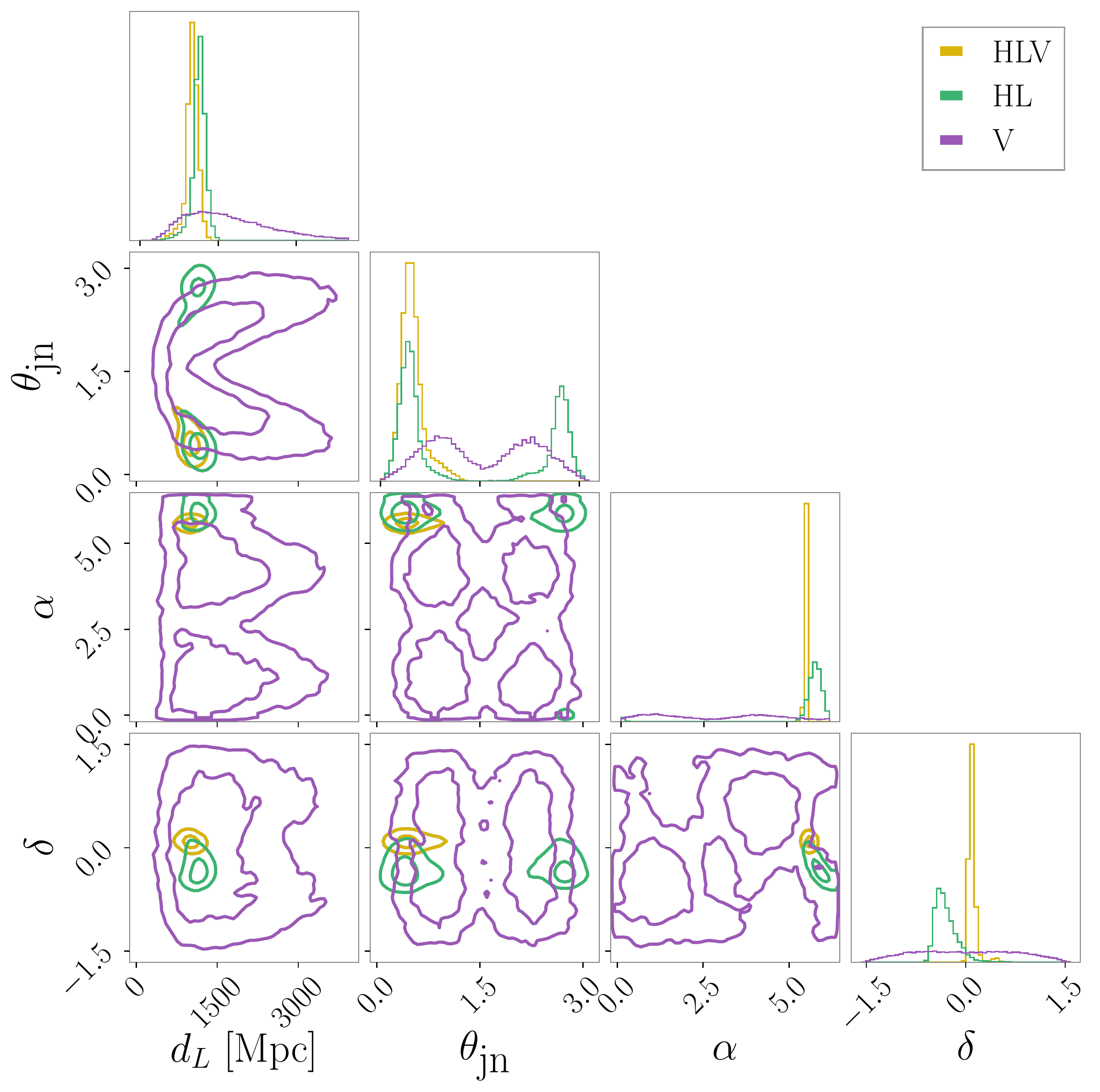}
    \caption{Similar to the right panel of Fig.~\ref{fig:intrinsic} but for select extrinsic parameters: luminosity distance $d_L$, angle between total angular momentum and line of sight $\theta_\textrm{jn}$, right ascension $\alpha$, and declination $\delta$. For reference, the median optimal SNR for each run is HLV: 27.6, HL: 26.9, V: 6.7.}
    \label{fig:extrinsic}
\end{figure}

The role of Virgo data on the inferred binary extrinsic parameters is explored in Fig.~\ref{fig:extrinsic}. In general, Virgo data have a larger influence on the extrinsic than the intrinsic parameters as the measured time and amplitude helps break existing degeneracies.
The extrinsic parameter posteriors show a large degree of overlap. The Virgo distance posterior does not rail against the upper prior cut off, suggesting that this detector does observe some excess power. The HL sky localization also overlaps with the Virgo-only one, though the latter is merely the antenna pattern of the detector that excludes the four Virgo ``blind spots."
We use the HL results to calculate the projected waveform in Virgo and calculate the 90\% lower limit on the signal SNR to be $4.2$. This suggests that given the LIGO data, Virgo should be observing a signal with at least that SNR at the 90\% level.

\begin{figure*}
    \centering
    \includegraphics[width=\linewidth]{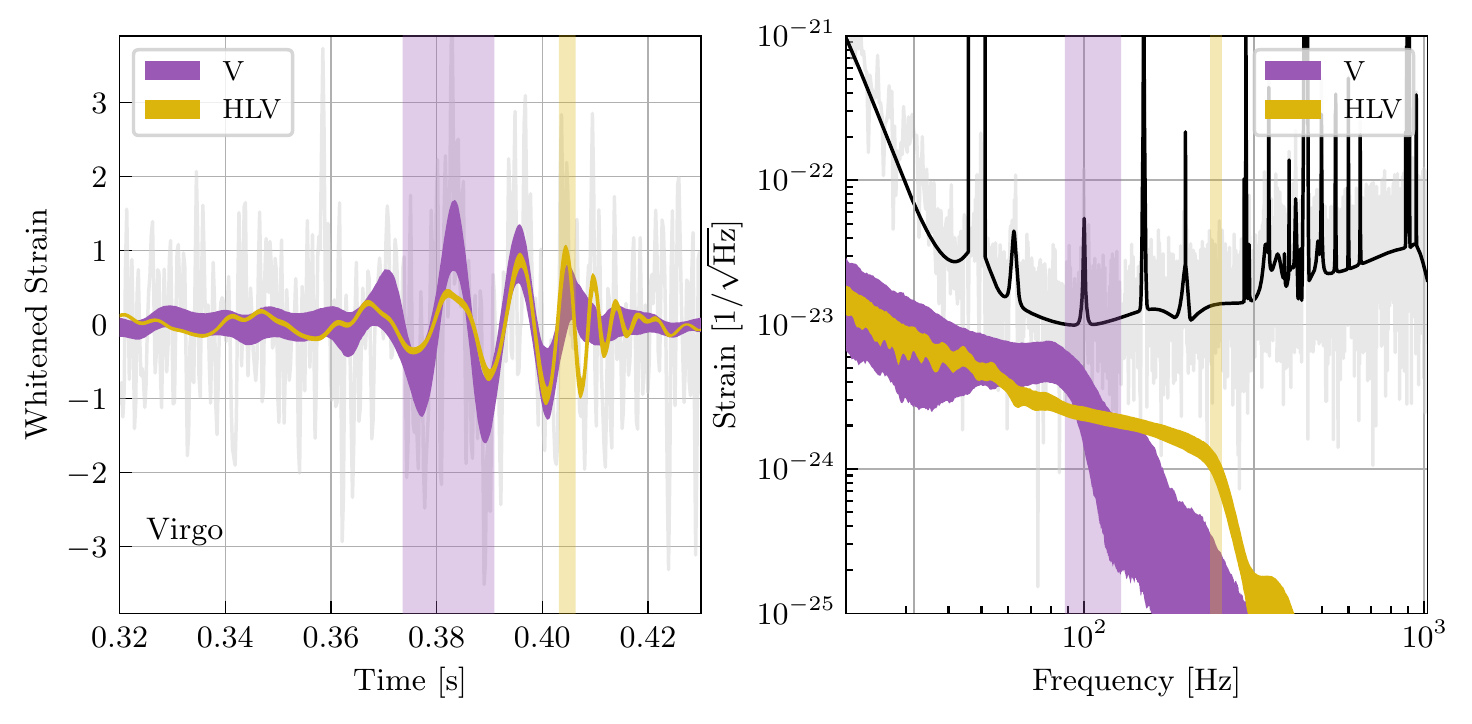}
    \caption{90\% credible intervals for the whitened time-domain reconstruction (left) and spectrum (right) of the signal in Virgo from a Virgo-only (purple; V) and a full 3-detector (yellow; HLV) analysis, see Table~\ref{tab:Bilbyrunsettings} for analysis settings. The data are shown in gray and the noise PSD in black. The time on the left plot is relative to GPS 1264316116. The high value of the PSD at $\sim50$\,Hz was imposed due to miscalibration of the relevant data~\cite{gwtc3}. Vertical shaded regions at each panel correspond to the 90\% credible intervals of the merger time (left; defined as the time of peak strain amplitude) and merger frequency (right; approximated via the dominant ringdown mode frequency as computed with {\tt qnm}~\cite{Stein:2019mop}, merger remnant properties were computed with {\tt surfinBH}~\cite{Varma:2018aht}). The Virgo data point to a heavier binary that merges $\sim 20$ms earlier than the full 3-detector results that are dominated by the LIGO detectors.}
    \label{fig:Vreconstruction}
\end{figure*}

In order to track down the cause of the discrepancy in the inferred mass parameters, we examine the Virgo strain data directly. Figure~\ref{fig:Vreconstruction} shows the whitened time-domain reconstruction (left panel) and the spectrum (right panel) of the signal in Virgo from a Virgo-only and a full 3-detector analysis. Compared to Figs.~\ref{fig:intrinsic} and~\ref{fig:extrinsic}, here we only consider a 3-detector analysis as the reconstructed signal in Virgo inferred from solely LIGO data would not be phase-coherent with the data, and thus would be uninformative. Given the higher signal SNR in the two LIGO detectors, the signal reconstruction morphology in Virgo is driven by them, as evident from the intrinsic parameter posteriors from the right panel of Fig.~\ref{fig:intrinsic}. 

The two reconstructions in Fig.~\ref{fig:Vreconstruction} are morphologically distinct. The 3-detector inferred signal is dominated by the LIGO data and resembles a typical ``chirp" with increasing amplitude and frequency. This signal is, however, inconsistent with the Virgo data as it underpredicts the strain at $t\sim$0.382\,s in the left panel. The Virgo-only inferred signal matches the data better by instead placing the merger at earlier times to capture the increased strain at $t\sim$0.382\,s as shown by the shaded vertical region denoting the merger time. Rather than a ``chirp," the signal is dominated by the subsequent ringdown phase with an amplitude that decreases slowly over $\sim$2 cycles. As also concluded from the inferred masses in Fig.~\ref{fig:intrinsic}, the Virgo data point to a heavier binary with lower ringdown frequency (vertical regions in the right panel).

\begin{figure*}
    \centering
    \includegraphics[width=\textwidth]{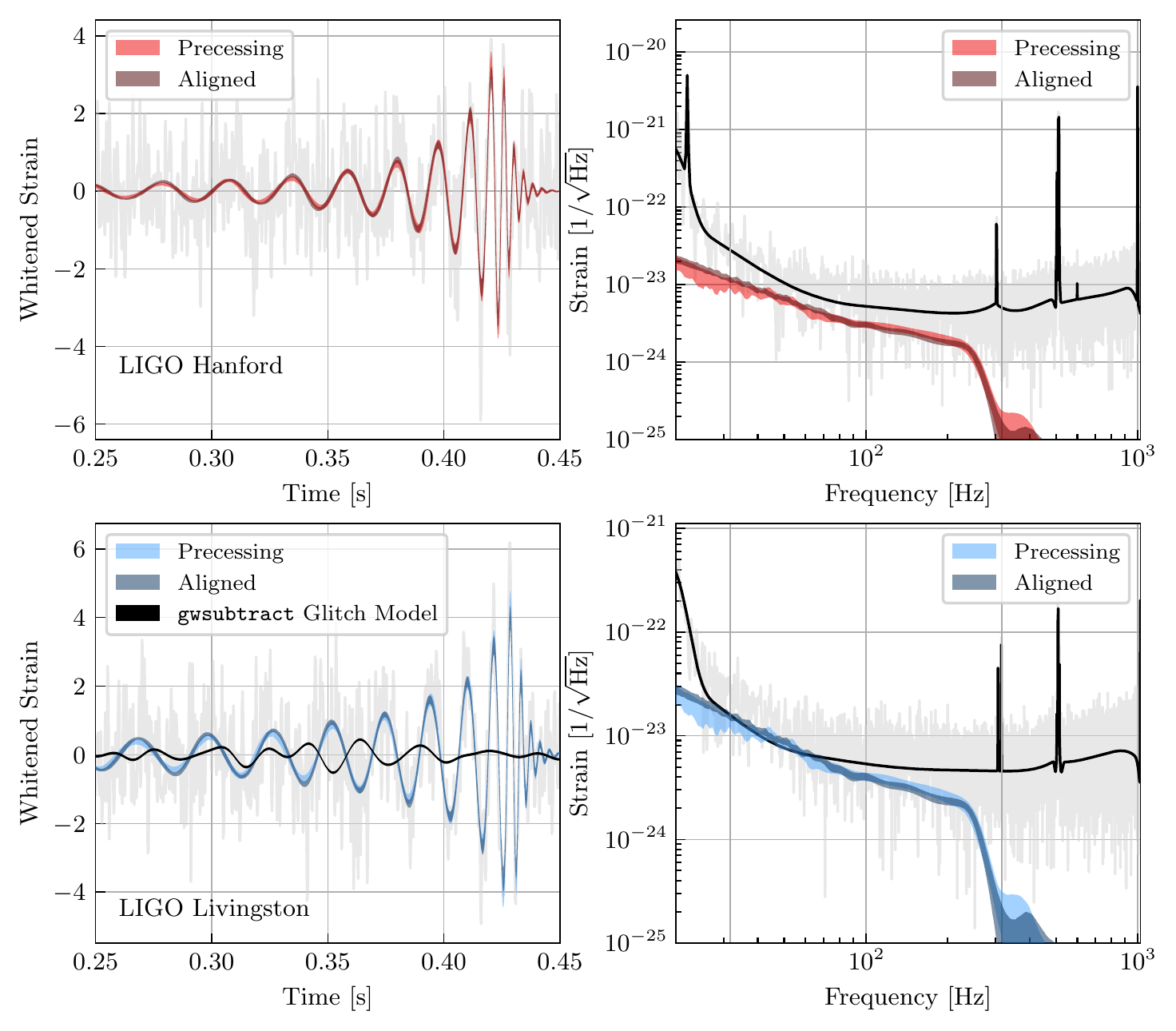}
    \caption{Whitened time-domain reconstruction (left) and spectrum (right) of GW200129 in LIGO Hanford (top) and LIGO Livingston (bottom). Shaded regions show the 90\% credible intervals for the signal using a spin-precessing (light blue and red) and a spin-aligned (dark blue and red) analysis based on {\tt NRSur7dq4}, see Table~\ref{tab:Bilbyrunsettings} for run settings. In gray we show the analyzed data where the {\tt gwsubtract} estimate for the glitch (black line) has already been subtracted. The black line in the right panels is the noise PSD. The glitch overlaps with the part of the inferred signal where the spin-aligned amplitude is on average larger than the spin-precessing one. }
    \label{fig:HLreconstruction}
\end{figure*}

Despite these large inconsistencies, the issues with the Virgo data do not affect our main goal, which is identifying the origin of the evidence for spin-precession. In order to avoid further ambiguities for the remainder of this section we restrict to data from the two LIGO detectors unless otherwise noted. In Fig.~\ref{fig:intrinsic} we concluded that LIGO Livingston alone drives this measurement and here we attempt to further zero in on the data that support spin-precession by comparing results from a spin-precessing and a spin-aligned analysis with {\tt NRSur7dq4}, see App.~\ref{appendix:bilby} for details. Figure~\ref{fig:HLreconstruction} shows the whitened time-domain reconstruction (left panel) and the spectrum (right panel) in LIGO Hanford (top) and LIGO Livingston (bottom). The two reconstructions remain phase-coherent, however there are some differences in the inferred amplitudes, with the spin-aligned amplitude being slightly larger at $\sim $30--50\,Hz and slightly smaller for $\gtrsim 100$\,Hz.  Comparison to the estimate for the glitch that was subtracted from the data based on information from auxiliary channels with {\tt gwsubtract} shows that the glitch overlaps with the part of the signal where the spin-precessing amplitude is smaller than the spin-aligned one. The glitch subtraction and data quality issues are therefore related to the evidence for spin-precession.

\begin{figure}
    \centering
    \includegraphics[scale=0.45]{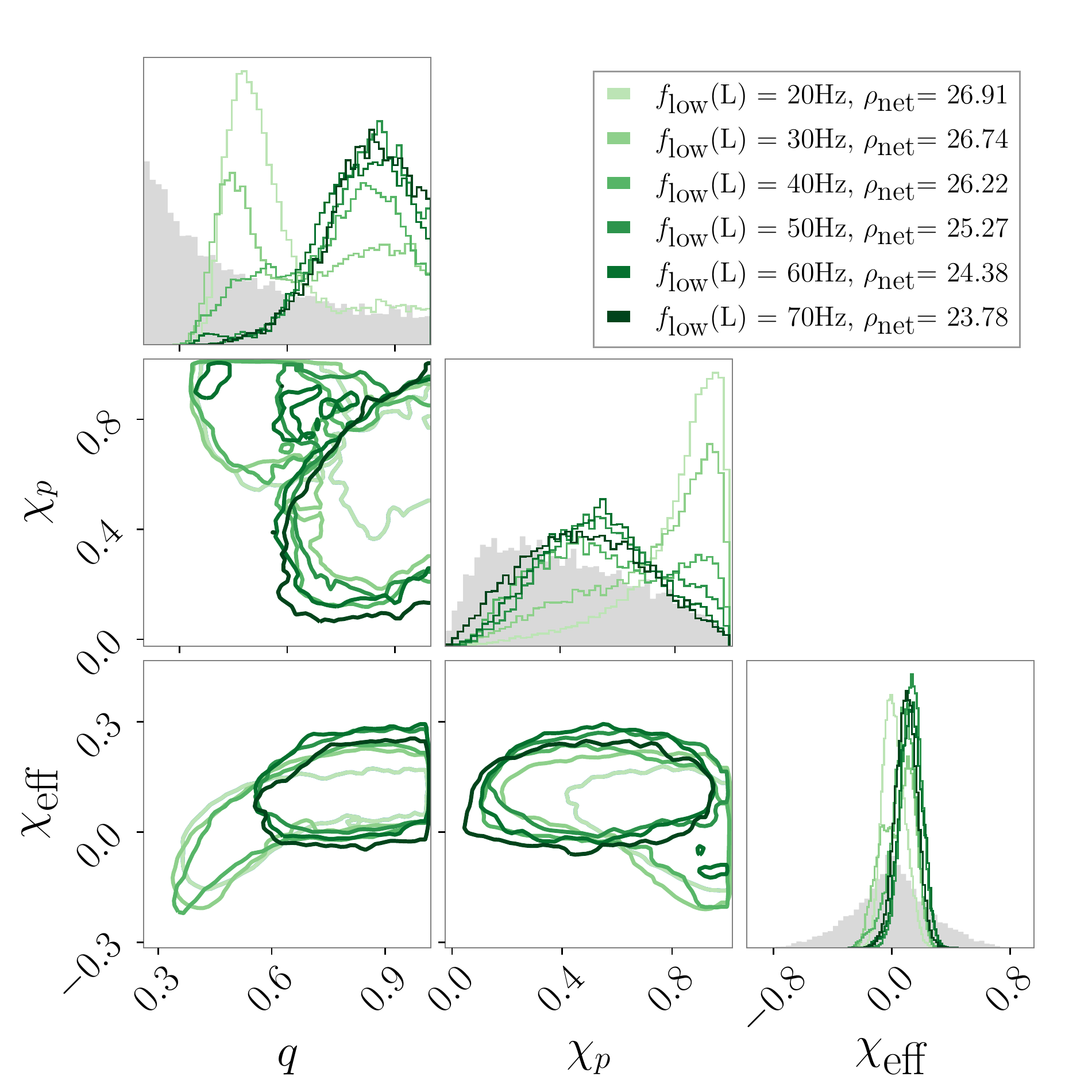}
    \caption{One- and two-dimensional marginalized posterior for the mass ratio $q$, the precession parameter $\chip$, and the effective spin parameter $\chi_\text{eff}$ for analyses using a progressively increasing low frequency cutoff in LIGO Livingston but all the LIGO Hanford data, see Table~\ref{tab:Bilbyrunsettings} for details. The median network SNR for each value of the frequency cutoff is given in the legend. Contours represent 90\% credible regions and the prior is shaded in gray. As the glitch-affected data are removed from the analysis, the posterior approaches that of an equal-mass binary and becomes uninformative about $\chip$. This behavior does not immediately indicate data quality issues and we only use this increasing-$\flow(L)$ analysis to isolate the data which contribute the evidence of spin-precession when compared to the rest of the data to within 20--50\,Hz.}
    \label{fig:HLflow}
\end{figure}

We confirm that the low-frequency data in LIGO Livingston (in relation to the rest of the data) are the sole source of the evidence for spin-precession, by carrying out analyses with a progressively increasing low frequency cutoff in LIGO Livingston only, while leaving the LIGO Hanford data intact. Figure~\ref{fig:HLflow} shows the effect on the posterior for $\chip$, $q$, and $\chieff$. When we use the full data bandwidth, $\flow(L)=20$\,Hz, we find that $q$ and $\chip$ are correlated and their two-dimensional posterior appears similar to the combination of the individual-detector posteriors from Fig.~\ref{fig:intrinsic}. However, as the low frequency cutoff in LIGO Livingston is increased and the data affected by the glitch are removed, the posterior progressively becomes more consistent with an equal-mass binary and $\chip$ approaches its prior. By $\flow(L)=50$\,Hz, $\chip$ is similar to its prior and further increasing $\flow(L)$ has a marginal effect. This confirms that \emph{given all the other data}, the LIGO Livingston data in 20--50\,Hz drive the inference for spin-precession.

The signal network SNR (i.e., the SNR in both detectors added in quadrature) is given in the legend for each value of the low frequency cutoff. By $\flow(L)=50$\,Hz where all evidence for spin-precession has been eliminated, the SNR reduction is only 1.5 units, suggesting that the large majority of the signal is consistent with a non-precessing origin. This might also suggest that $\chip$ inference is not degraded solely due to loss of SNR, as the latter is very small.
The $\chieff$ posterior is generally only minimally affected, with a small shift to higher values driven by the $q-\chieff$ correlation~\cite{Cutler:1994ys}. We have verified that these conclusions are robust against re-including the Virgo data (using their full bandwidth). 

\begin{figure*}
    \centering
    \includegraphics[width=\linewidth]{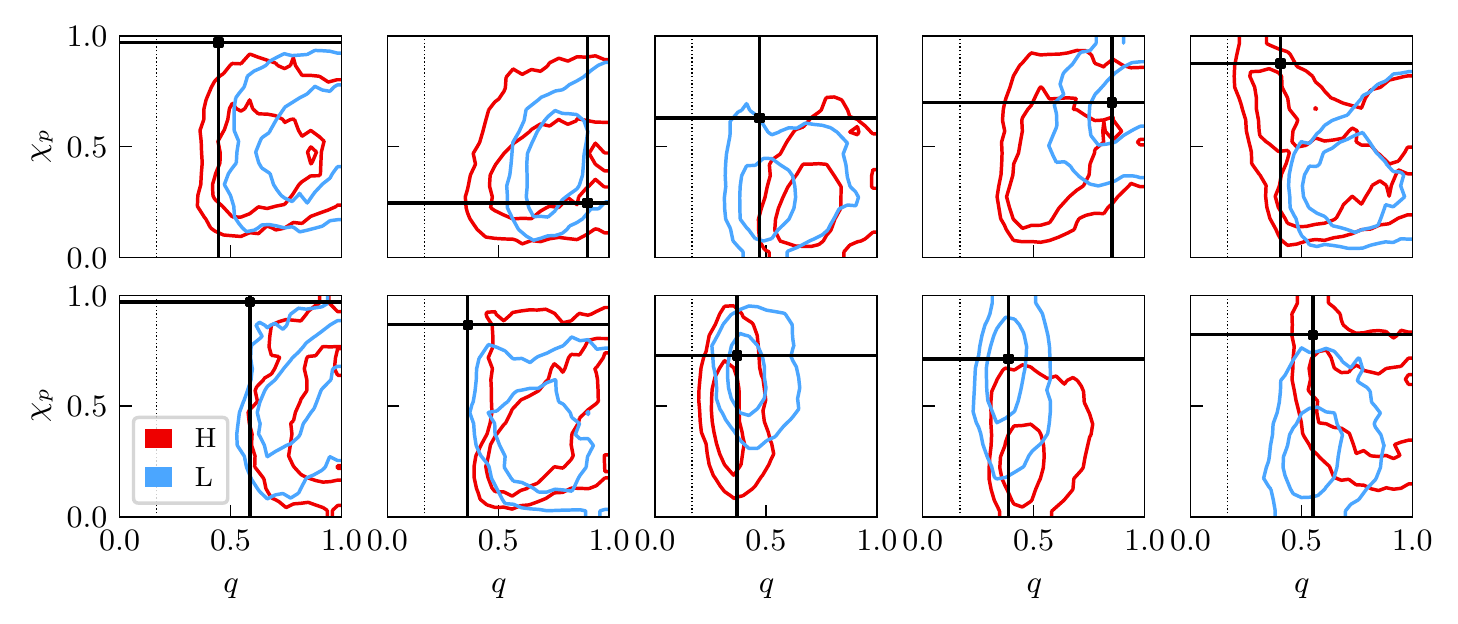}
    \caption{90\% contours for the two-dimensional marginalized posteriors for the mass ratio $q$ and the precessing parameter $\chip$ obtained from analyzing data from each LIGO detector separately for $10$ simulated signals. The signal parameters are drawn from the posterior for GW200129 when using LIGO Livingston data only and true values are indicated by black lines. Due to the spin priors disfavoring large $\chip$, the injected value is outside the two-dimensional 90\% contour in some cases. We only encounter an inconsistency between LIGO Hanford (red; H) and LIGO Livingston (blue; L) as observed for GW200129 in Fig.~\ref{fig:intrinsic} in $\mathcal{O}(5/100)$ injections.}
    \label{fig:wallpaper}
\end{figure*}

The above analysis is \emph{not} on its own an indication of data quality issues in LIGO Livingston, but we now turn to an observation that might be more problematic: the $q-\chip$ inconsistency between LIGO Hanford and LIGO Livingston identified in Fig.~\ref{fig:intrinsic}. In order to examine whether such an effect could arise from the different Gaussian noise realizations in each detector, we consider simulated signals. We use $100$ posterior samples obtained from analyzing solely the LIGO Livingston data, make simulated data that include a noise realization with the same noise PSDs as GW200129, and analyze data from the two LIGO detectors separately. 
To quantify the degree to which the LIGO Hanford and LIGO Livingston posteriors overlap, we compute the Bayes factor for overlapping posterior distributions relative to if the two distributions do not overlap~\cite{Haris:2018vmn, Hannuksela:2019kle},
\begin{equation}
    \mathcal{B}^{\textrm{overlapping}}_{\textrm{not overlapping}} = \iint \textrm{d}\chip \textrm{d}q\, \frac{p_L(\chip,q|d)p_H(\chip,q|d)}{\pi(\chip, q)},
\end{equation}
where we compute the overlap within the $q-\chip$ plane, $p_L(\chip,q|d)$ and $p_H(\chip,q|d)$ are the LIGO Livingston and LIGO Hanford posteriors, and $\pi(\chip, q)$ is the prior. While evaluating this quantity is subject to sizeable sampling uncertainty for events where the two distributions are more distinct (i.e., the case of GW200129), we find $\mathcal{O}(5/100)$ injections have a similar overlap as GW200129 (Fig.~\ref{fig:intrinsic}).
Figure~\ref{fig:wallpaper} shows a selection of $q-\chip$ posteriors for 10 injections as inferred from each detector separately.
The posteriors typically overlap, though they are shifted with respect to each other as expected from the different noise realizations.

We conclude that the evidence for spin-precession originates exclusively from the LIGO Livingston data that overlapped with a glitch. This causes an inconsistency between the LIGO Hanford and LIGO Livingston that we typically do not encounter in simulated signals in pure Gaussian noise. This inconsistency suggests that there might be residual data quality issues in LIGO Livingston that were not fully resolved by the original glitch subtraction. Though inconsequential for the spin-precession investigation, we also identify severe data quality issues in Virgo. Before returning to the investigation of spin-precession, we first examine the Virgo data in detail in Sec.~\ref{sec:glitchV} and argue that they should be removed from subsequent analyses. We reprise the spin-precession investigations in Sec.~\ref{sec:glitchL}.

%%%%%%%%%%%%%%%%%%%%%%%%%%%%%%%%%%%%%%%%%%%
\section{Data quality issues: Virgo}
\label{sec:glitchV}
%%%%%%%%%%%%%%%%%%%%%%%%%%%%%%%%%%%%%%%%%%%

\begin{figure}
    \centering
    \includegraphics[width=\linewidth]{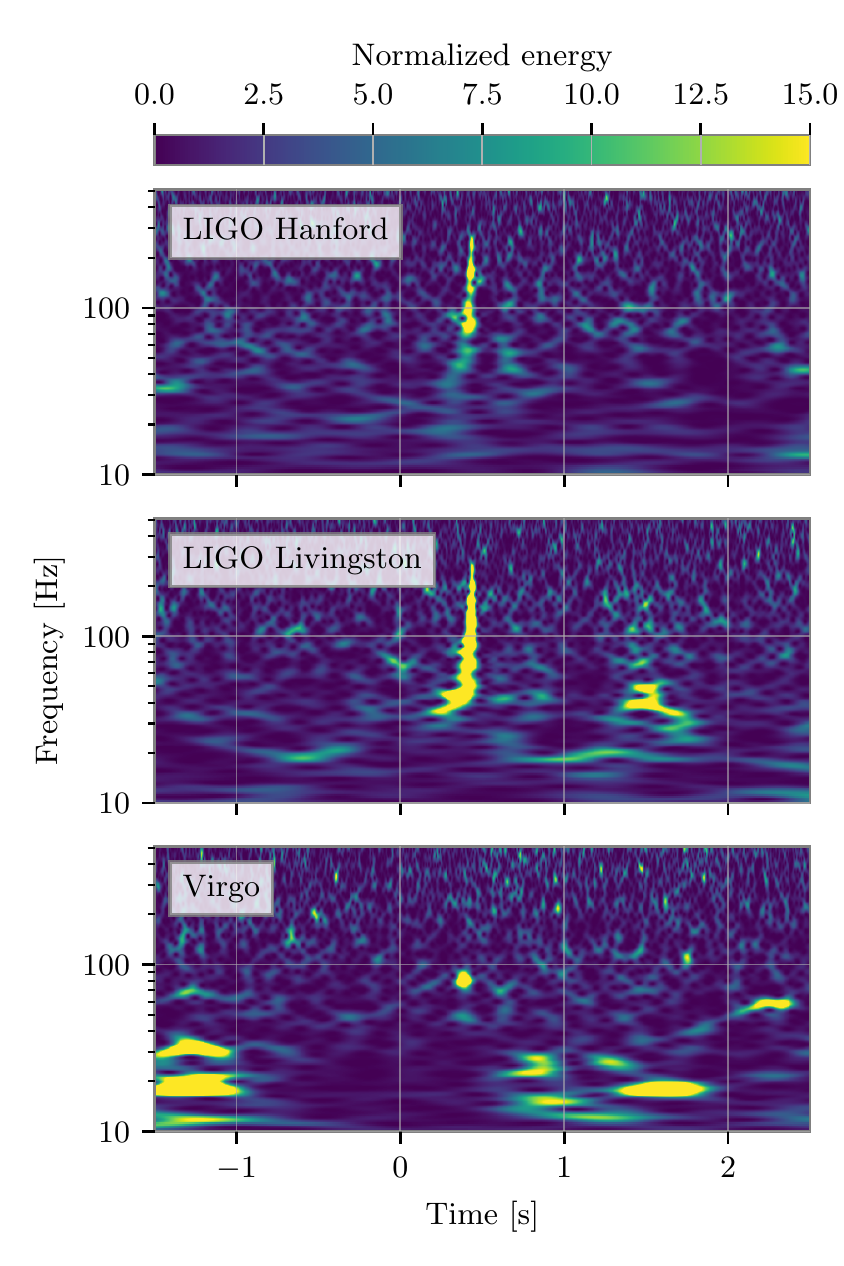}
    \caption{Spectrogram of the data in each detector, plotted using plotted using the Q-transform~\cite{Chatterji:2004qg,2021SoftX..1300657M}. Listed times are with respect to GPS 1264316116. Besides the clear chirp morphology in LIGO, there is visible excess power $\sim1$\,s after the signal in LIGO Livingston. Virgo demonstrates a high rate of excess power, though most is due to scattered light and concentrated at frequencies $<30$\,Hz. The excess power in Virgo that is coincident with GW200129 does not have a chirp morphology.}
    \label{fig:spectrogram}
\end{figure}

Having established that the Virgo trigger is coincident but not fully coherent with the triggers in the two LIGO detectors, we explore potential reasons for this discrepancy. Figure~\ref{fig:spectrogram} shows a spectrogram of the data in each detector centered around the time of the event. A clear chirp morphology is visible in the LIGO detectors but not in Virgo, though this might also be due to the low SNR of the Virgo trigger. Within a few seconds of the trigger, however, a number of other glitches are also present in Virgo, mostly assigned to scattered light. We estimate the SNR of the Virgo trigger without assuming it is a CBC signal (i.e., without using a CBC model) through Omicron~\cite{Robinet:2020lbf} and {\tt BayesWave} using its glitch model that fits the data with sine-Gaussian wavelets, see Table~\ref{tab:BWrunsettings} for run settings\footnote{=The {\tt BayesWave} analyses described here does not concurrently marginalize over the PSD uncertainty.}. The former finds a matched-filter Omicron SNR\footnote{The SNR reported by Omicron is normalized so that the expectation value of the SNR is 0, rather than $\sqrt{2}$~\cite{Robinet:2020lbf}. To highlight this difference, we use the phrase ``Omicron SNR" whenever a reported result uses this normalization.} of $7.0$, while the latter finds an optimal SNR of $7.3$ for the median glitch reconstruction.

Given the prevalence of glitches, the first option is that the Virgo trigger is actually a detector glitch that happened to coincide with a signal in the LIGO detectors. To estimate the probability that such a coincidence could happen by chance, we consider the glitch rate in Virgo. In O3, the median rate of glitches in Virgo was $1.11/$min, with significant variation versus time~\cite{gwtc3}. When we consider the hour of data around the event, the rate of glitches with Omicron SNR $>6.5$ is $10.2/$min. Most of the glitches in Virgo at this time are due to scattered light~\cite{Accadia:2010zzb,Longo:2020onu,Longo:2021avq,Virgo:2022fxr,LIGO:2020zwl}. While Fig.~\ref{fig:spectrogram} shows that there are scattered light glitches in the Virgo data near the time of GW200129, the excess power from these glitches are concentrated at frequencies $<30$\,Hz. To account for the excess power corresponding to GW200129 in Virgo, there must be a different type of glitch present in the data. The rate of glitches at frequencies similar to the signal is much lower; using data from 4 days around the event, the rate of glitches with frequency 60-120\,Hz is only $0.06/$hr. Given this rate, we calculate the probability that a glitch occurred in Virgo within a 0.06\,s window (roughly corresponding to twice the light-travel time between the LIGO detectors and Virgo) around a trigger in the LIGO detectors. We find that if glitches at any frequency are considered, the probability of coincidence per event is $\mathcal{O}(0.01)$, and if only glitches with similar frequencies are considered, the same probability is $\mathcal{O}(10^{-5})$.

Another option is that the Virgo trigger is a combination of a genuine signal and a detector glitch. We explore this possibility using {\tt BayesWave}~\cite{Cornish:2014kda,Littenberg:2014oda,Cornish:2020dwh} to simultaneously model a potential CBC signal that is coherent across the detector network and overlapping glitches that are incoherent~\cite{Chatziioannou:2021ezd,Hourihane:2022doe}. In this ``CBC+glitch" analysis, {\tt BayesWave} models the CBC signal with the {\tt IMRPhenomD} waveform~\cite{Husa:2015iqa,Khan:2015jqa} and glitches with sine-Gaussian wavelets. Details about the models and run settings are provided in App.~\ref{appendix:bayeswave}.
An important caveat here is that {\tt IMRPhenomD} does not include the effects of higher-order modes and spin-precession. A concern is, therefore, that the CBC model could fail to model precession-induced modulations in the signal amplitude and instead assign them to the glitch model. This precise scenario is tested in Hourihane \textit{et al.}~\cite{Hourihane:2022doe} where the analysis was shown to be robust against such systematics. Below we argue that the same is true here for the Virgo data, especially since they are consistent with a spin-aligned binary as shown in Fig.~\ref{fig:intrinsic}.

\begin{figure}
    \centering
    \includegraphics[width=\linewidth]{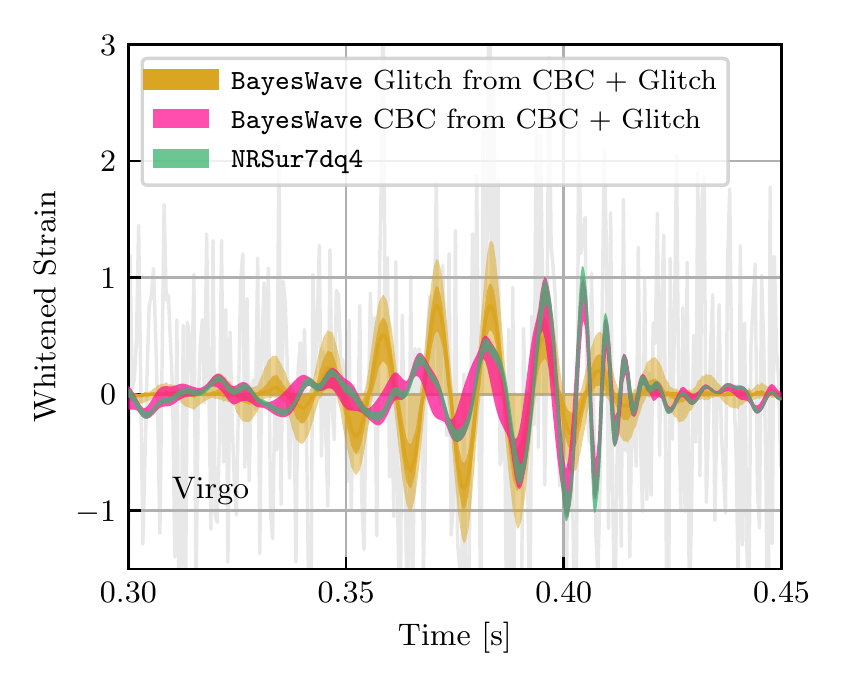}
    \caption{Whitened time-domain reconstruction of the signal in Virgo obtained after analysis of data from all three detectors relative to GPS 1264316116. Shaded regions correspond to 90\% and 50\% (where applicable) credible intervals. Green corresponds to the same 3-detector result obtained with {\tt NRSur7dq4} as Fig.~\ref{fig:Vreconstruction}, while pink and gold correspond to the CBC and glitch part of the ``CBC+glitch" analysis with {\tt BayesWave}. See Tables~\ref{tab:Bilbyrunsettings} and~\ref{tab:BWrunsettings} for run settings. The two CBC reconstructions largely overlap, suggesting that the lack of spin-precession in {\tt BayesWave}'s analysis does not affect the reconstruction considerably. A glitch overlapping with the signal is, however, recovered. }
    \label{fig:BWBilbyinV}
\end{figure}

Figure~\ref{fig:BWBilbyinV} compares {\tt BayesWave}'s reconstruction in Virgo with the one obtained with the {\tt NRSur7dq4} analysis from Fig.~\ref{fig:Vreconstruction} that ignores a potential glitch but models spin-precession and higher order modes. All results are obtained using data from all three detectors. The CBC reconstruction from {\tt BayesWave} with {\tt IMRPhenomD} is consistent with the one from {\tt NRSur7dq4} to within the 90\% credible level at all times. This is unsurprising given Fig.~\ref{fig:intrinsic} that shows that Virgo data are consistent with a spin-aligned BBH. Crucially, there is no noticeable difference between the two CBC reconstructions for times when the inferred glitch is the loudest. This suggests that the lack of higher-order modes and spin-precession in {\tt IMRPhenomD} does not lead to a noticeable difference in the signal reconstruction and could thus not account for the inferred glitch. 
The differences between the inferred signals using {\tt IMRPhenomD} and {\tt NRSur7dq4} are much smaller than the amount of incoherent power present in Virgo.
In fact, the glitch reconstruction is larger than the signal at the 50\% credible level, though not at the 90\% level. This result suggests that a potential explanation for the trigger in Virgo is a combination of a signal consistent with the one in the LIGO detectors and a glitch.

\begin{figure}
    \centering
    \includegraphics[width=\linewidth]{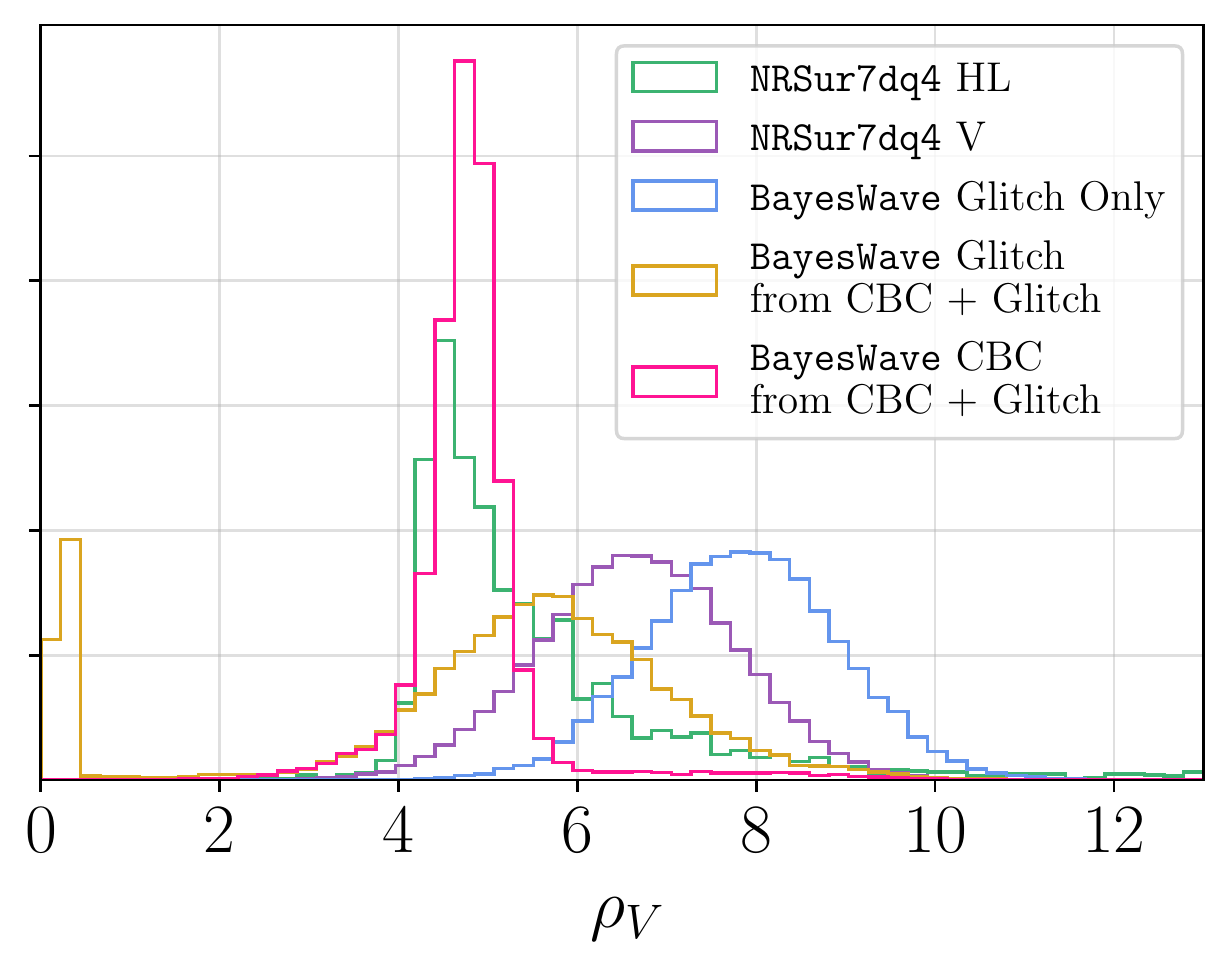}
    \caption{Comparison of optimal SNR estimates for Virgo from different analyses. In green is the posterior for the expected SNR in Virgo from just the LIGO data using the \texttt{NRSur7dq4} waveform (HL analysis of Fig.~\ref{fig:intrinsic}), while purple corresponds to the SNR from an analysis of the Virgo data only (V analysis of Fig.~\ref{fig:intrinsic}). The CBC and glitch SNR posterior from {\tt BayesWave}'s full ``CBC+glitch" model (Fig.~\ref{fig:BWBilbyinV}) are shown in pink and orange respectively. Part of the latter is consistent with zero, which corresponds to no glitch (as also seen from the 90\% credible interval in Fig.~\ref{fig:BWBilbyinV}). The SNR posterior from a ``glitchOnly" {\tt BayesWave} is shown in blue.}
    \label{fig:SNRV}
\end{figure}

Figure~\ref{fig:SNRV} summarizes the various SNR estimates for the excess power in Virgo. We plot an estimate of the SNR in Virgo suggested by LIGO data; in other words it is the SNR that is consistent with GW200129 as observed by LIGO. In comparison, we also show the SNR from a Virgo-only analysis and the SNR from {\tt BayesWave}'s ``glitchOnly" analysis that models the excess power with sine-Gaussian wavelets without the requirement that it is consistent with a CBC. The fact that the SNR inferred from HL data is smaller than the other two again suggests that the Virgo trigger is not consistent with the one seen by LIGO and contains additional power. {\tt BayesWave}'s ``CBC+glitch" analysis is able to separate the part of the trigger that is consistent with a CBC and recovers a CBC SNR that is consistent to the one inferred from LIGO only. The ``remaining" power is assigned to a glitch with SNR $\sim4.6$ (computed through the median {\tt BayesWave} glitch reconstruction).

Based on the glitch SNR calculated by the {\tt BayesWave} ``CBC+glitch" model, we revisit the probability of overlap with a signal based on the SNR distribution of Omicron triggers.
Since the lowest SNR recorded in Omicron analyses is 5.0, we fit the SNR distribution of glitches with Omicron SNR $>5.0$ with a power-law and extrapolate to SNR 4.6.
We find that the rate of glitches with frequencies similar to the one in Fig.~\ref{fig:BWBilbyinV} with SNR $>4.6$ is $0.31/$min and the probability of overlap with a signal in Virgo is $\mathcal{O}(10^{-3})$. Given the 60 events from GWTC-3 that were identified in Virgo during O3, the overall chance of at least one glitch of this SNR overlapping a signal is $\mathcal{O}(0.1)$.  

The above studies suggest that the most likely scenario is that the Virgo trigger consists of a signal and a glitch. However, due to the low SNR of both, this interpretation is subject to sizeable statistical uncertainties and we therefore do not attempt to make glitch-subtracted Virgo data. Such data would be extremely dependent on which glitch reconstruction we chose to subtract, for example the median or a fair draw from the {\tt BayesWave} glitch posterior. For these reasons and due to its low sensitivity, we do not include Virgo data in what follows.

%%%%%%%%%%%%%%%%%%%%%%%%%%%%%%%%%%%%%%%%%%%
\section{Data quality issues: LIGO Livingston}
\label{sec:glitchL}
%%%%%%%%%%%%%%%%%%%%%%%%%%%%%%%%%%%%%%%%%%%

The data quality issues in LIGO Livingston were identified and mitigated in GWTC-3~\cite{gwtc3} through use of information from auxiliary channels~\cite{Davis:2018yrz,Davis:2022ird} and the {\tt gwsubtract} pipeline as also described in App.~\ref{appendix:glitchsubtraction}. The comparison of Figs.~\ref{fig:intrinsic} and~\ref{fig:wallpaper}, however, suggest that residual data quality issues might remain, as the two LIGO detectors result in inconsistent inferred $q-\chip$ parameters beyond what is expected from typical Gaussian noise fluctuations.
Here we revisit the LIGO Livingston glitch with {\tt BayesWave} and again model both the CBC and potential glitches. This analysis offers a point of comparison to {\tt gwsubtract} as it uses solely strain data to infer the glitch instead of auxiliary channels. Additionally, {\tt BayesWave} computes a posterior for the glitch, rather than a single point estimate, and thus allows us to explore the statistical uncertainty of the glitch mitigation.
In all analyses involving {\tt BayesWave} we use the original LIGO Livingston data without any of the data mitigation described in App.~\ref{appendix:glitchsubtraction}. 

\begin{figure}
    \centering
    \includegraphics[width=\linewidth]{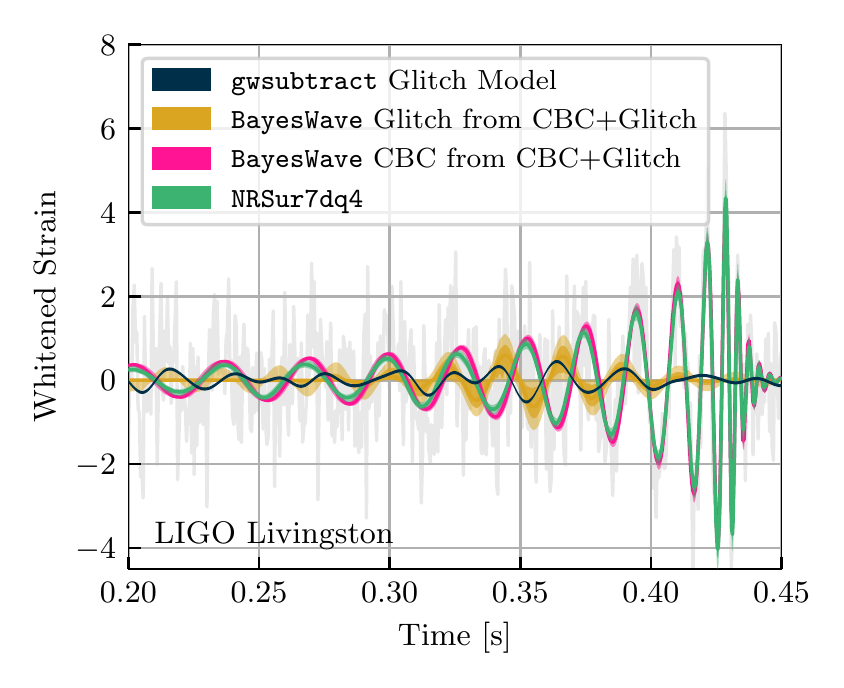}
    \caption{Whitened time-domain reconstruction of the data in LIGO Livingston obtained after analysis of data from the two LIGO detectors. Shaded regions correspond to 90\% and 50\% (where applicable) credible intervals and gray gives the original data without any glitch mitigation. Green corresponds to the same 2-detector result obtained with {\tt NRSur7dq4} as Fig.~\ref{fig:HLreconstruction}, while pink and gold correspond to the CBC and glitch part of the joint ``CBC+glitch" analysis with {\tt BayesWave}. The black line shows an estimate for the glitch obtained through auxiliary channels. All analyses use only LIGO data.}
    \label{fig:BWBilbyinL}
\end{figure}

Figure~\ref{fig:BWBilbyinL} shows {\tt BayesWave}'s CBC and glitch reconstructions in LIGO Livingston compared to the one based on the {\tt NRSur7dq4} (from glitch-mitigated data) and the glitch model computed with {\tt gwsubstract}. All analyses use data from the two LIGO detectors only. Unsurprisingly, now, the CBC reconstructions based on {\tt IMRPhenomD} and {\tt NRSur7dq4} do not fully overlap around t=$0.3$\,s, though they are consistent during the signal merger phase. This is expected from the fact that LIGO Livingston supports spin-precession as well as Fig.~\ref{fig:HLreconstruction}. However, this difference is \emph{smaller} than the statistical uncertainty in the inferred glitch from {\tt BayesWave} (yellow) and well as differences between the {\tt BayesWave} and the {\tt gwsubtract} glitch estimates. This suggests that even though the {\tt BayesWave} glitch estimate might be affected by the lack of spin-precession in its CBC model, this effect is smaller than the glitch uncertainty. 

We also model the signal as a superposition of coherent wavelets in addition to the incoherent glitch wavelets using {\tt BayesWave}~\cite{Cornish:2014kda,Littenberg:2014oda,Cornish:2020dwh}. This approach has been previously utilized for glitch subtraction~\cite{gwtc3}. However, we do not recover strong evidence for a glitch overlapping the signal in LIGO Livingston when running with this ``signal+glitch" analysis. The ``signal+glitch'' analysis attempts to describe both the signal and the glitch with wavelets and hence it is significantly less sensitive than the ``CBC+glitch'' model. In the data of interest, both the signal and the glitch whitened amplitudes are $\sim1\sigma$ and as such they are difficult to separate using coherent and incoherent wavelets. Given that we know (based on the auxiliary channel data) that there is some non-Gaussian noise in LIGO Livingston, we find that the ``signal+glitch" analysis is not sensitive enough for our data.

\begin{figure*}
    \centering
    \includegraphics[width=\linewidth]{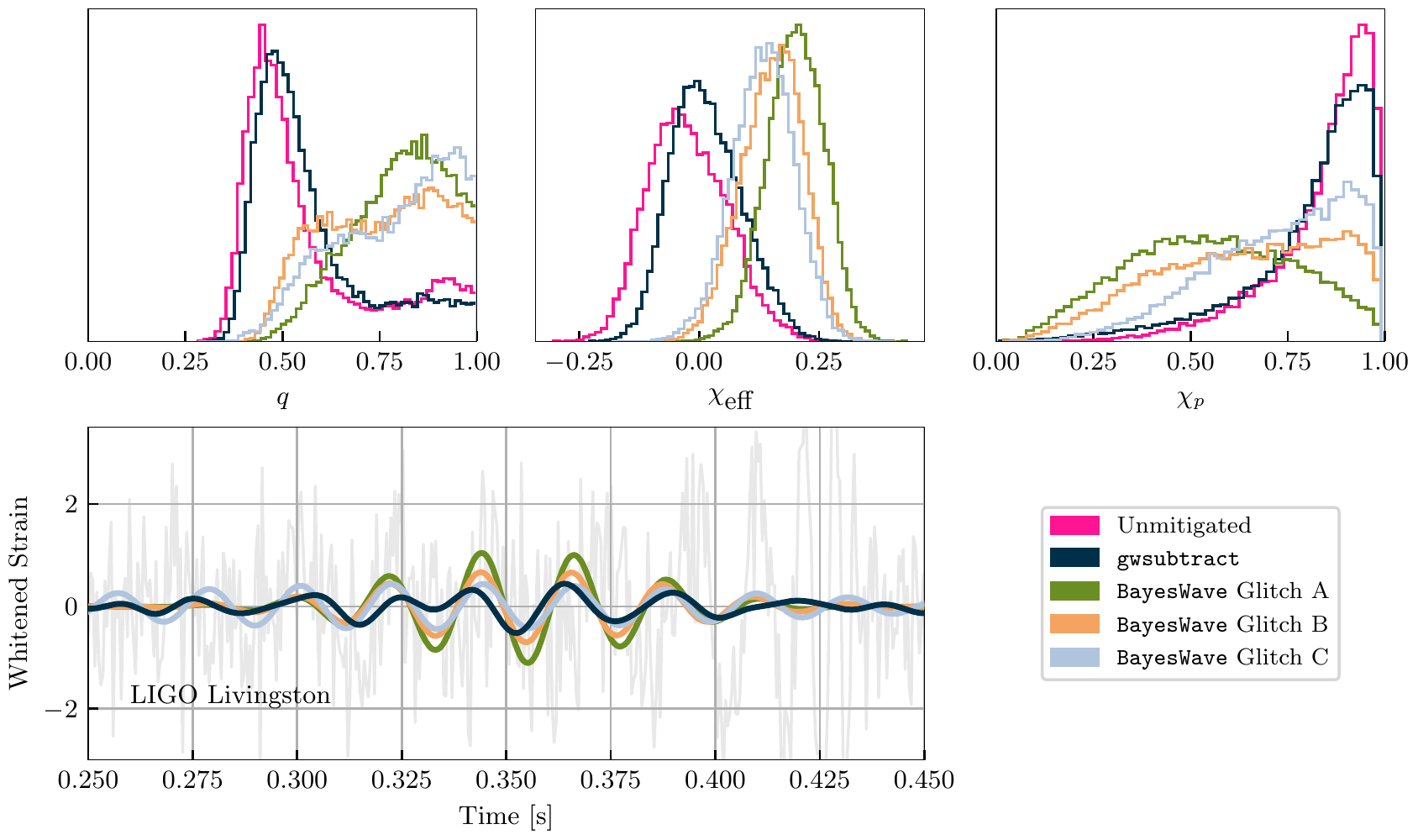}
    \caption{Bottom: Whitened, time domain reconstructions of various glitch reconstructions subtracted from LIGO Livingston data. The green line corresponds to the glitch reconstruction obtained from auxiliary data using {\tt gwsubtract}. The rest are glitch posterior draws from the {\tt BayesWave} ``CBC+Glitch" analysis on HL unmitigated data. Top: Marginalized posterior distributions corresponding to parameter estimation performed with the {\tt NRSur7dq4} waveform model on HL data where each respective glitch realization was subtracted from LIGO Livingston (same colors). Pink corresponds to the original data without any glitch subtraction.
    Larger glitch reconstruction amplitudes roughly lead to less informative $\chip$ posteriors and eliminate the $q-\chip$ inconsistency between LIGO Hanford and LIGO Livingston.}
    \label{fig:GlitchDraws}
\end{figure*}

The large statistical uncertainty in the glitch reconstruction (yellow bands in Fig.~\ref{fig:BWBilbyinL}) implies that the difference between the spin-precessing and non-precessing interpretation of GW200129 cannot be reliably resolved. To confirm this, we select three random samples from the glitch posterior of Fig.~\ref{fig:BWBilbyinL}, subtract them from the unmitigated LIGO Livingston data, and repeat the parameter estimation analysis with {\tt NRSur7dq4}. The {\tt BayesWave} glitch-subtracted frames and associated {\tt NRSur7dq4} parameter estimation results are available in~\citep{data_release}. For reference, we also analyze the original unmitigated data (no glitch subtraction whatsoever). Figure~\ref{fig:GlitchDraws} confirms that the spin-precession evidence depends sensitively on the glitch subtraction. The original unmitigated data and the {\tt gwsubtract} subtraction yield the largest evidence for spin-precession, but this is reduced -or completely eliminated- with different realizations of the {\tt BayesWave} glitch model. In general, larger glitch amplitudes lead to less support for spin-precession, suggesting that the evidence for spin-precession is increased when the glitch is \emph{undersubtracted}. 

\begin{figure*}
    \centering
    \includegraphics[width=\linewidth]{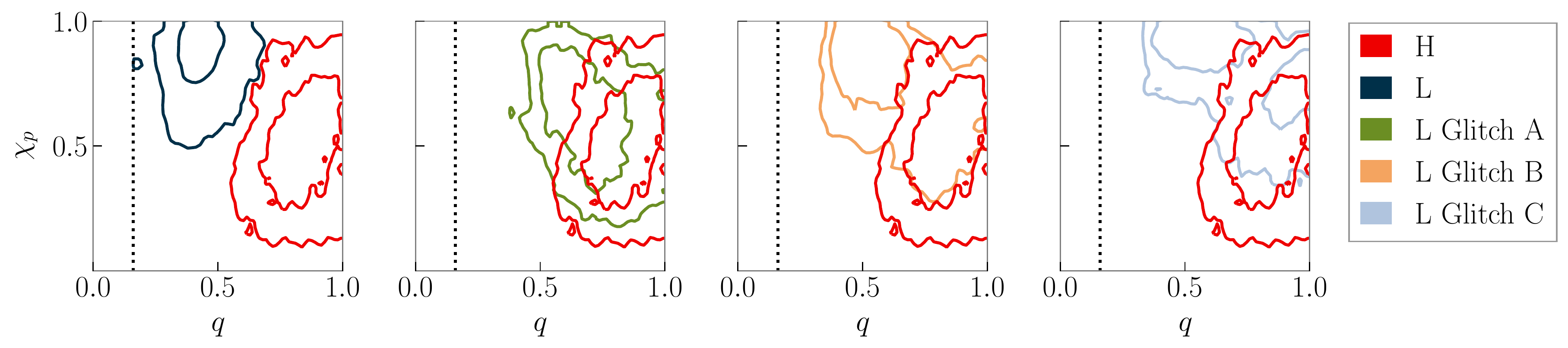}
    \caption{Two-dimensional posterior distributions for $\chip$ and $q$ (50\% and 90\% contours) from single-detector parameter estimation runs. The far left panel shows the same tension as the LIGO Hanford and LIGO Livingston data plotted in Fig.~\ref{fig:intrinsic} when using the {\tt gwsubtract} estimate for the glitch. Subsequent figures show inferred posterior distributions using data where the same three different {\tt BayesWave} glitch models as Fig.~\ref{fig:GlitchDraws} have been subtracted. These results show less tension between the two posterior distributions.}
    \label{fig:Lglitches}
\end{figure*}

Figure~\ref{fig:Lglitches} compares the corresponding $q-\chip$ posterior inferred from LIGO Hanford and LIGO Livingston separately under each different estimate for the glitch. Each of the $3$ {\tt BayesWave} glitch draws results in single-detector posteriors that fully overlap, thus resolving the inconsistency seen in $q-\chip$ when using the {\tt gwsubtract} glitch estimate. 
Due to the lack of spin-precession modeling in the ``CBC+glitch" analysis of Fig.~\ref{fig:BWBilbyinL}, however, we cannot definitively conclude that any one of the new glitch-subtracted results is preferable. The $3$ {\tt BayeWave} glitch draws result in different levels of support for spin-precession, it is therefore possible that GW200129 is still consistent with a spin-precessing system.
We do conclude, though, that the evidence for spin-precession is contingent upon the large statistical uncertainty of the glitch subtraction.

As a further check of whether the lack of spin-precession in {\tt BayesWave}'s CBC model could severely bias a potential glitch recovery, we revisit the $10$ simulated signals from Fig.~\ref{fig:wallpaper} and analyze them with the ``CBC+glitch" model. These signals are consistent with GW200129 as inferred from LIGO Livingston data only, and thus exhibit the largest amount of spin-precession consistent with the signal. In all cases we find that the glitch part of the ``CBC+glitch" model has median and 50\% credible intervals that are consistent with zero at all times. This again confirms that the differences between the spin-precessing and the spin-aligned inferred signals in Fig.~\ref{fig:BWBilbyinL} is smaller than the uncertainty in the glitch. This tests suggests that the glitch model is not strongly biased by the lack of spin-precession, however it does not preclude small biases (within the glitch statistical uncertainty); it is therefore necessary but not sufficient.

\begin{figure}
    \centering
    \includegraphics[width=\linewidth]{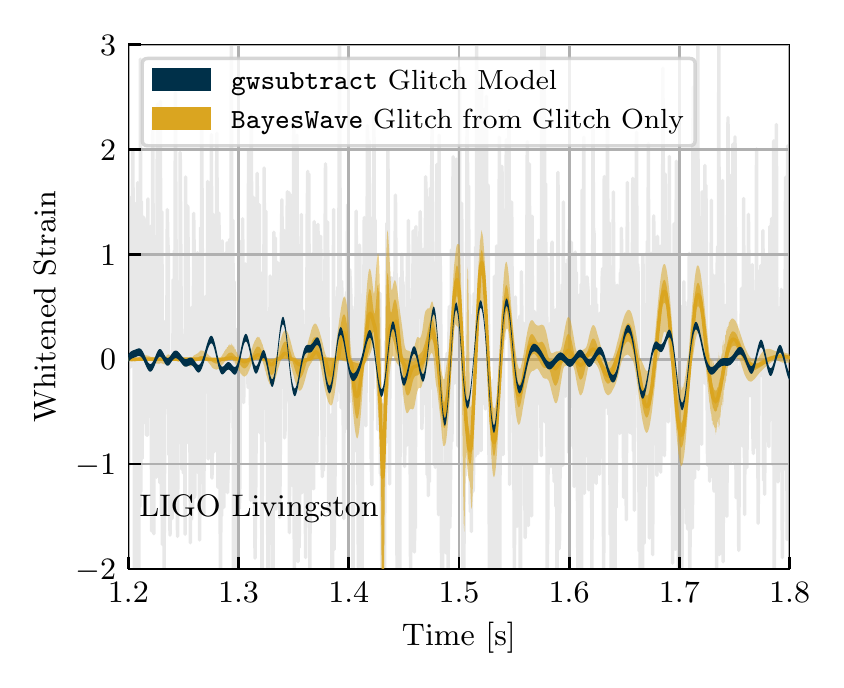}
    \caption{Comparison between the two glitch reconstruction and subtraction methods for a glitch in LIGO Livingston $\sim1$\,s after GW200129, see the middle panel of Fig.~\ref{fig:spectrogram}. We plot the original data with no glitch mitigation (grey), the glitch reconstruction obtained from auxiliary channels with 90\% confidence intervals (black), and the 50\% and 90\% credible intervals for the glitch obtained with {\tt BayesWave} that uses only the strain data (gold).}
    \label{fig:LLOGlitchCompare}
\end{figure}

As a final point of comparison between {\tt BayesWave}'s glitch reconstruction that is based on strain data and the {\tt gwsubtract} glitch reconstruction based on auxiliary channels, we consider a \emph{different} glitch in LIGO Livingston approximately $1$s after the signal, see Fig.~\ref{fig:spectrogram}. Studying this glitch offers the advantage of direct comparison of the two glitch reconstruction methods without contamination from the CBC signal and uncertainties about its modeling. We analyze the original data with no previous glitch mitigation around that glitch using {\tt BayesWave}'s glitch model and plot the results in Fig.~\ref{fig:LLOGlitchCompare}. For the {\tt gwsubtract} reconstruction we also include 90\% confidence intervals, as described in App.~\ref{appendix:glitchsubtraction}.

The two estimates of the glitch are broadly similar but they do not always overlap within their uncertainties. The main disagreement comes from the sharp data ``spike" at $t=1.43$\,s that is missed by {\tt gwsubtract}, but recovered by {\tt BayesWave}. The reason is that the the maximum frequency considered by {\tt gwsubtract} was 128\,Hz and thus cannot capture such a sharp noise feature~\cite{Davis:2022ird}. Away from the ``spike," the two glitch estimates are approximately phase-coherent. On average {\tt BayesWave} recovers a larger glitch amplitude as the {\tt gwsubtract} result typically falls on {\tt BayesWave}'s lower 90\% credible level. 

Figures~\ref{fig:BWBilbyinL} and~\ref{fig:LLOGlitchCompare} broadly suggest that {\tt BayesWave} recovers a higher-amplitude glitch. Figure~\ref{fig:GlitchDraws} shows that the evidence for spin-precession is indeed reduced, the LIGO Hanford-LIGO Livingston inconsistency is alleviated (Fig.~\ref{fig:Lglitches}), and the LIGO Livingston data become more consistent across low and high frequencies (Fig.~\ref{fig:HLflow}) if the glitch was originally undersubtracted. 
However, due to the low SNR of the glitch and other systematic uncertainties it is not straightforward to select a ``preferred" set of glitch-subtracted data. All studies, however, indicate that the statistical uncertainty of the glitch amplitude is larger than the difference between the inferred spin-precessing and spin-aligned signals.  

%%%%%%%%%%%%%%%%%%%%%%%%%%%%%%%%%%%%%%%%%%%
\section{Conclusions}
\label{sec:conclusions}
%%%%%%%%%%%%%%%%%%%%%%%%%%%%%%%%%%%%%%%%%%%

Though it might be possible to infer the presence of spin-precession and large spins in heavy BBHs, our investigations suggest that in the case of GW200129 any such evidence is contaminated by data quality issues in the LIGO Livingston detector. In agreement with~\cite{Hannam:2021pit} we find that the evidence for spin-precession originates exclusively from data from that detector. However, we go beyond this and also demonstrate the following.
\begin{enumerate}
    \item The evidence for spin-precession in LIGO Livingston is localized in the 20--50\,Hz band in comparison to the rest of the data, precisely where the glitch overlapped the signal. Excluding this frequency range from the analysis, we find that GW200129 is consistent with an equal-mass BBH with an uninformative $\chip$ posterior; it is thus similar to the majority of BBH detections~\cite{LIGOScientific:2018jsj,LIGOScientific:2020kqk,LIGOScientific:2021psn}. However, the fact that there is no evidence for spin-precession if $\flow(L)>50$\,Hz is not on its own cause for concern as it might be due to Gaussian noise fluctuations or the precise precessional dynamics of the system.
    \item LIGO Hanford is not only uninformative about spin-precession (which again could be due to Gaussian noise fluctuations or the lower signal SNR in that detector), but it also yields an \emph{inconsistent} $q-\chip$ posterior compared to LIGO Livingston. Using simulated signals, we find that the latter, i.e., the $q-\chip$ inconsistency, is larger than $\mathcal{O}(95\%)$ of results expected from Gaussian noise fluctuations.
    \item Given the LIGO Livingston glitch's low SNR, the statistical uncertainty in modeling it is \emph{larger} than the difference between a spin-precessing and a non-precessing analysis for GW200129. Inferring the presence of spin-precession requires reliably resolving this difference, something challenging as we found by using different realizations of the glitch model from the {\tt BayesWave} glitch posterior. Crucially, any evidence for spin-precession in GW200129 depends sensitively on the glitch model and priors employed.
    \item Given the large statistical uncertainty in modeling the glitch, evidence for systematic differences between {\tt BayesWave} and {\tt gwsubtract} that use strain and auxiliary data respectively is tentative. However, the {\tt BayesWave} estimate typically predicts a larger glitch amplitude, which would reduce the evidence for spin-precession and alleviate the tension between LIGO Hanford and LIGO Livingston. Additionally, we do not recover any support for a glitch when injecting spin-precessing signals from the LIGO Livingston-only posterior distribution into Gaussian noise. This indicates that {\tt BayesWave} is unlikely to be strongly biasing the glitch recovery due to its lack of spin-precession. 
\end{enumerate}
Overall, given the uncertainty surrounding the LIGO Livingston glitch mitigation, we cannot conclude that the source of GW200129 was spin-precessing. We do not conclude the opposite either, however. Though we obtain tentative evidence that the glitch was undersubtracted, we can at present not estimate how much it was undersubtracted by due to large statistical and potential systematic uncertainties. It is possible that some evidence for spin-precession remains, albeit reduced given the glitch statistical uncertainty.

In addition, we verify that this uncertainty in the glitch modeling is larger than uncertainty induced by detector calibration. We repeat select analyses in Appendix~\ref{appendix:bilby} and confirm that the inclusion of uncertainty in the calibration of the gravitational-wave detectors negligibly impacts the spin-precession inference, as expected. Indeed, the glitch impacts the data at a level comparable to the signal strain, c.f., Fig.~\ref{fig:BWBilbyinL}, whereas the calibration uncertainty within $20$ to $70$\,Hz is only $\sim5\%$ in amplitude and $5^\circ$ in phase~\cite{Sun:2020wke}. Therefore, the glitch in LIGO Livingston's data dominates over uncertainties about the data calibration.

Though not critical to the discussion and evidence for spin-precession, we also identified data quality issues in Virgo. The inconsistency between Virgo and the LIGO detectors is in fact more severe than the one between the two LIGO detectors, however the Virgo data do not influence the overall signal interpretation due to the low signal SNR in Virgo. Nonetheless, we argue that the most likely explanation is that the Virgo data contain both the GW200129 signal and a glitch.

These conclusions are obtained with {\tt NRSur7dq4}, which is expected to be the more reliable waveform model including spin-precession and higher-order modes in this region of the parameter space~\cite{Varma:2019csw,Hannam:2021pit}. We repeated select analyses with {\tt IMRPhenomXPHM} which also favored a spin-precessing interpretation for GW200129~\cite{gwtc3}. We found largely consistent but not identical results between {\tt NRSur7dq4} and {\tt IMRPhenomXPHM}, suggesting that there are additional systematic differences between the two waveform models. Appendix~\ref{app:systematics} shows some example results. Nonetheless, our results are directly comparable to the ones of~\cite{Hannam:2021pit,Varma:2022pld} as they were obtained with the same waveform model.

Our analysis suggests that extra caution is needed when attempting to infer the role of subdominant physical effects in the detected GW signals, for example spin-precession or eccentricity. Low-mass signals are dominated by a long inspiral phase that in principle allows for the detection of multiple spin-precession cycles or eccentricity-induced modulations. However, the majority of detected events, such as GW200129, have high masses and are dominated by the merger phase. The subtlety of the effect of interest and the lack of analytical understanding might make inference susceptible not only to waveform systematics, but also (as argued in this study) potential small data quality issues.

Indeed, Fig.~\ref{fig:GlitchDraws} shows that a difference in the glitch amplitude of $<0.5\sigma$ can make the difference between an uninformative $\chip$ posterior and one that strongly favors spin-precession. 
This also demonstrates that low-SNR glitches are capable of biasing inference of these subtle physical effects. Low-SNR departures from Gaussian noise have been commonly observed by statistical tests of the residual power present in the strain data after subtracting the best-fit waveform of events~\cite{LIGOScientific:2019fpa,LIGOScientific:2020tif,LIGOScientific:2021sio}. 
If indeed such low-SNR glitches are prevalent, they might be individually indistinguishable from Gaussian noise fluctuations. Potential ways to safeguard our analyses and conclusions against them are (i) the detector and frequency band consistency checks performed here, (ii) extending the {\tt BayesWave} ``CBC+glitch" analysis to account for spin-precession and eccentricity while carefully accounting for the impact of glitch modeling and priors especially for low SNR glitches, (iii) and modeling insight on the morphology of subtle physical effects of interest such as spin-precession and eccentricity in relation to common detector glitch types.

\acknowledgements

We thank Aaron Zimmerman, Eric Thrane, Paul Lasky, Hui Tong, Geraint Pratten, Mark Hannam, Charlie Hoy, Jonathan Thompson, Steven Fairhurst, Vivien Raymond, Max Isi, and Colm Talbot for useful discussions. We also thank Vijay Varma for providing a version of {\tt LALSuite} optimized for running {\tt NRSur7dq4}, as well as suggestions for some of our configuration settings.
This research has made use of data, software and/or web tools obtained from the Gravitational Wave Open Science Center (https://www.gw-openscience.org), a service of LIGO Laboratory, the LIGO Scientific Collaboration and the Virgo Collaboration.
Virgo is funded by the French Centre National de Recherche Scientifique (CNRS), the Italian Istituto Nazionale della Fisica Nucleare (INFN) and the Dutch Nikhef, with contributions by Polish and Hungarian institutes.
This material is based upon work supported by NSF's LIGO Laboratory which is a major facility fully funded by the National Science Foundation.
The authors are grateful for computational resources provided by the LIGO Laboratory and supported by NSF Grants PHY-0757058 and PHY-0823459.
This research utilized the OzStar Supercomputing Facility at Swinburne Unersity of Technology. The OzStar facility is partially funded by the Astronomy National Collaborative Research Infrastructure Strategy (NCRIS) allocation provided by the Australian Government.  
This research was enabled in part by computing resources provided by Simon Fraser University and the Digital Research Alliance of Canada (alliancecan.ca).
SH and KC were supported by NSF Grant PHY-2110111.
Software: {\tt gwpy}~\cite{2021SoftX..1300657M}, {\tt matplotlib}~\cite{Hunter:2007}, {\tt numpy}~\cite{harris2020array}, {\tt pandas}~\cite{reback2020pandas}, {\tt scipy}~\cite{2020SciPy-NMeth}, {\tt qnm}~\cite{Stein:2019mop}, {\tt surfinBH}~\cite{Varma:2018aht}, {\tt Bilby}~\cite{bilbygit}, {\tt LALSuite}~\cite{lalsuite}, {\tt BayesWave}~\cite{bayeswave}.

\appendix

%%%%%%%%%%%%%%%%%%%%%%%%%%%%%%%%%%%%%%%%%%%%%%%
\section{Analysis details}
\label{appendix:analysis}
%%%%%%%%%%%%%%%%%%%%%%%%%%%%%%%%%%%%%%%%%%%%%%%

In this Appendix we provide details and settings for the analyses presented in the main text. All data are obtained via the GW Open Science Center~\cite{LIGOScientific:2019lzm}. Throughout we use geometric units, $G=c=1$.

%%%%%%%%%%%%%%%%%%%%%%%%%%%%%%%%%%%%%%%%%%%%%%%
\subsection{Detection and Glitch-subtracted data}
\label{appendix:glitchsubtraction}
%%%%%%%%%%%%%%%%%%%%%%%%%%%%%%%%%%%%%%%%%%%%%%%

GW200129 was identified in low latency~\cite{GCN26926} by GstLAL~\cite{Messick:2016aqy,Hanna:2019ezx}, cWB~\cite{Klimenko:2015ypf}, PyCBC Live~\cite{Nitz:2018rgo,DalCanton:2020vpm}, MBTAOnline~\cite{Adams:2015ulm}, and SPIIR~\cite{Chu:2020pjv}.
The quoted false alarm rate of the signal in low latency was approximately 1 in $10^{23}$ years, making this an unambiguous detection. Below we recap the detection and glitch mitigation process from~\cite{gwtc3}.

Multiple data quality issues were identified in the data surrounding GW200129.
As a part of the rapid response procedures, scattered light noise~\cite{Accadia_2010,Virgo:2022fxr} was identified in the Virgo data, as seen in Fig.~\ref{fig:spectrogram} in the frequency range 10--60\,Hz. 
These glitches did not overlap the signal, and no mitigation steps were taken with the Virgo data.
During offline investigations of the LIGO Livingston data quality, a malfunction of the 45 MHz electro-optic modulator system~\cite{LIGOScientific:2016gtq} was found to have caused numerous glitches in the days surrounding GW200129. 
To help search pipelines differentiate these types from glitches, a data quality flag was generated for this noise source~\cite{T2100045}. 
These data quality vetoes are used by some pipelines to veto any candidates identified during the data quality flag time segments~\cite{LIGO:2021ppb}.
The glitches from the electro-optic modulator system directly overlapped GW200129, meaning that the time of the signal overlapped the time of the data quality flag. 

Although clearly an astrophysical signal, the data quality issues present in LIGO Livingston introduced additional complexities into the estimation of the significance of this signal~\cite{gwtc3}. 
Due to the data quality veto, the signal was not identified in LIGO Livingston by the PyCBC~\cite{Nitz_2017,Davies:2020tsx} MBTA~\cite{Aubin:2020goo}, and cWB~\cite{Klimenko:2015ypf} pipelines. 
PyCBC was still able to identify GW200129 as a LIGO Hanford -- Virgo detection, but the signal was not identified by MBTA due to the high SNR in LIGO Hanford and cWB due to post-production cuts.
The GstLAL~\cite{Sachdev:2019vvd,Cannon:2020qnf} analysis did not incorporate data quality vetoes in its O3 analyses and was therefore able to identify the signal in all three detectors. 

The excess power from the glitch directly overlapping GW200129 in LIGO Livingston was subtracted before estimation of the signal's source properties~\cite{gwtc3,Davis:2022ird} using the \texttt{gwsubtract} algorithm~\cite{Davis:2018yrz}.
This method relies on an auxiliary sensor at LIGO Livingston that also witnesses glitches present in the strain data. 
The transfer function between the sensor and the strain data channel is measured using a long stretch of data by calculating the inner product of the two time series with a high frequency resolution and then averaging the measured value at nearby frequencies to produce a transfer function with lower frequency resolution~\cite{Allen:1999wh}. 
This transfer function is convolved with the auxiliary channel time series to estimate the contribution of this particular noise source to the strain data. 
Therefore, the effectiveness of this subtraction method is limited by the accuracy of the auxiliary sensor and the transfer function estimate.
This tool was previously used for broadband noise subtraction with the O2 LIGO dataset~\cite{Davis:2018yrz}, but this was the first time it was used for targeted glitch subtraction.
Additional details about the use of \texttt{gwsubtract} for the GW200129 glitch subtraction can be found in Davis \textit{et al.}~\cite{Davis:2022ird}.

The \texttt{gwsubtract} glitch model does not include a corresponding interval that accounts for all sources of statistical errors as is done by \texttt{BayesWave}. 
However, a confidence interval based on only uncertainties due to random correlations between the auxiliary channel and the strain data can be computed. 
For the GW200129 glitch model, this interval is $\pm 0.022$ in the whitened strain data~\cite{Davis:2022ird}. 
Additional systematic uncertainties due to time variation in the measured transfer function and effectiveness of the chosen auxiliary channel are expected to be present but are not quantified. 
The relative size of these uncertainties is dependent on the specific noise source that is being modeled and chosen auxiliary channel.

%%%%%%%%%%%%%%%%%%%%%%%%%%%%%%%%%%%%%%%%%%%%%%%
\subsection{{\tt Bilby} parameter estimation analyses}
\label{appendix:bilby}
%%%%%%%%%%%%%%%%%%%%%%%%%%%%%%%%%%%%%%%%%%%%%%%

%
\begin{table*}[]
\begin{tabular}{c @{\quad\quad} c @{\quad\quad}c @{\quad}c @{\quad}c}
 Figure(s) & Waveform Model &  Detector Network  & Glitch mitigation & $\flow$ (Hz) \\
\hline
\ref{fig:intrinsic}, \ref{fig:Lglitches} & {\tt NRSur7dq4} &  H  & {\tt gwsubtract} & 20\\
\ref{fig:intrinsic}, \ref{fig:Lglitches} & {\tt NRSur7dq4} &  L & {\tt gwsubtract} & 20\\
\ref{fig:intrinsic}, \ref{fig:extrinsic}, \ref{fig:Vreconstruction} & {\tt NRSur7dq4} & V & {\tt gwsubtract} & 20\\
\ref{fig:intrinsic}, \ref{fig:extrinsic}, \ref{fig:Vreconstruction}, \ref{fig:BWBilbyinV} & {\tt NRSur7dq4} & HLV & {\tt gwsubtract} & 20\\
\ref{fig:intrinsic}, \ref{fig:extrinsic}, \ref{fig:HLreconstruction}, \ref{fig:BWBilbyinL}, \ref{fig:GlitchDraws}, \ref{fig:waveforms} & {\tt NRSur7dq4} & HL & {\tt gwsubtract} & 20\\
\ref{fig:HLreconstruction} & {\tt NRSur7dq4} spin-aligned & HL & {\tt gwsubtract} & 20\\
\multirow{2}{*}{\ref{fig:HLflow}} & \multirow{2}{*}{{\tt NRSur7dq4}} & \multirow{2}{*}{HL}  & \multirow{2}{*}{{\tt gwsubtract}} & \{20,30,40,50,60,70\} in L,\\
&&&& 20 in H\\
\ref{fig:GlitchDraws} & {\tt NRSur7dq4} &  HL  & No mitigation & 20\\
\ref{fig:GlitchDraws} & {\tt NRSur7dq4} &  HL  & {\tt BayesWave} fair draws & 20\\
\ref{fig:Lglitches} & {\tt NRSur7dq4} 
& L  & {\tt BayesWave} fair draws & 20\\
\ref{fig:waveforms} & {\tt IMRPhenomXPHM} & HL & {\tt gwsubtract} & 20\\
\hline
\hline
\end{tabular}
\caption{Table of {\tt Bilby} runs and settings. All analyses use 4\,s of data, and a sampling rate of 4096\,Hz. Columns correspond to the main text figures each analysis appears in, the waveform model, the detector network used (H: LIGO Hanford, L: LIGO Livingston, V: Virgo), the type of glitch mitigation in LIGO Livingston, and the low frequency cutoff of the analysis. Figure~\ref{fig:wallpaper} also presents results for a set of 10 injections drawn from the LIGO Livingston only posterior distribution with $\flow(L) = 20\,$Hz. These analyses use the same settings as above with $\flow(L) = 20\,$Hz.}
\label{tab:Bilbyrunsettings}
\end{table*}

Quasicircular BBHs are characterized by 15 parameters, divided into 8 intrinsic and 7 extrinsic parameters. 
Each component BH has source frame mass $m_i^{s}$, $i\in\{1,2\}$. 
In the main text we mainly use the corresponding detector frame (redshifted) masses $m_i = (1 + z)m_i^{s}$, where $z$ is the redshift, as we are interested in investigating data quality issues and detector frame quantities better relate to the signal as observed.
Each component BH also has dimensionless spin vector $\vec{\chi}_i$, and $\chi_i$ is the magnitude of this vector. 
We also use parameter combinations that are useful in various contexts: total mass $M = m_1 + m_2$, mass ratio $q = m_2/m_1<1$, chirp mass $\Mc= (m_1 m_2)^{3/5}(m_1 + m_2)^{-1/5}$~\cite{Poisson:1995ef, Blanchet:1995ez, Finn:1992xs}, effective orbit-aligned spin parameter~\cite{Racine:2008qv,Santamaria:2010yb,Ajith:2009bn}
\begin{equation}
\chieff = \frac{\vec{\chi}_1\cdot\vec{L} + q \vec{\chi}_2\cdot\vec{L}}{1 + q}\,,
\end{equation}
where $\vec{L}$ is the Newtonian orbital angular momentum, and effective precession spin parameter~\cite{Hannam:2013oca, Schmidt:2014iyl} %
\begin{equation}
    \chip = \text{max}\left(\chi_{1\perp}, q\chi_{2\perp}\frac{3q+4}{4q+3}\right)\,,
\end{equation}
where $\chi_{1\perp}$ is the $\vec{\chi}_i$ component that is perpendicular to $\vec{L}$.
The remaining parameters are observer dependent, and hence referred to as extrinsic. 
The right ascension $\alpha$ and declination $\delta$ designate the location of the source in the sky, while the luminosity distance to the source is $d_L$. The angle between total angular momentum and the observer's line of sight is $\theta_{jn}$; for systems without perpendicular spins it reduces to the inclination $\iota$, the angle between the orbital angular momentum and observer's line of sight.
The time of coalescence $t_c$ is the geocenter coalescence time of the binary. The phase of the signal $\phi$ is defined at a given reference frequency, and the polarization angle $\psi$ completes the geometric description of the sources position and orientation relative to us; neither of these are used directly in this work. 

Parameter estimation results are obtained with {\tt parallel Bilby}~\cite{Ashton:2018jfp,Romero-Shaw:2020owr, Smith:2019ucc} using the nested sampler, {\tt Dynesty}~\cite{Speagle2020dynestyAD}.
The numerical relativity surrogate, {\tt NRSur7dq4}~\cite{Varma:2019csw}, is used for all main results due to its accuracy over the regime of highly precessing signals. Its space of validity is limited by the availability of numerical simulations~\cite{Boyle:2019kee} to $q>1/4$ and component spin magnitudes $\chi<0.8$, though it maintains reasonable accuracy when extrapolated to $q>1/6$ and $\chi<1$~\cite{Varma:2019csw}.

The majority of our analyses use the publicly released strain data, including the aforementioned glitch subtraction in LIGO Livingston~\cite{Davis:2022ird}, and noise power spectral densities (PSDs)~\cite{gwtc3}. 
The exception to the publicly released data was the construction of glitch-subtracted strain data using {\tt BayesWave} for LIGO Livingston, as discussed in Sec.~\ref{sec:glitchL}. 
We do not incorporate the impact of uncertainty about the detector calibration as the SNR of the signal is far below the anticipated regime where calibration uncertainty is non-negligible~\cite{Vitale:2011wu, Payne:2020myg, Vitale:2020gvb, Essick:2022vzl}. Furthermore, we confirm that including marginalization of calibration uncertainty does not qualitatively change the recovered posterior distributions or our main conclusions by also directly repeating select runs.

As is done in GWTC-3~\cite{gwtc3}, we choose a prior that is uniform in detector frame component masses, while sampling in chirp mass and mass ratio. The mass ratio prior bounds are 1/6 and 1, where we utilize the extrapolation region of {\tt NRSur7dq4}. Since {\tt NRSur7dq4} is trained against numerical relativity simulations which typically have a short duration, only a limited number of cycles are captured before coalescence. With a reduced signal model duration, our analysis is restricted to heavier systems so that the model has content spanning the frequencies analyzed (20\,Hz and above). We therefore enforce an additional constraint on the total detector-frame mass to be greater than $60\,M_\odot$. We verify that our posteriors reside comfortably above this lower bound. 
The luminosity distance prior is chosen to be uniform in comoving volume. 
The prior distribution on the sky location is isotropic with a uniform distribution on the polarization angle. 
Finally, for most analyses, the prior on the spin distributions is isotropic in orientation and uniform in spin magnitude up to $\chi = 0.99$. 
For the spin-aligned analyses, a prior is chosen on the aligned spin to mimic an isotropic and uniform spin magnitude prior. 
These settings and data are utilized in conjunction with differing GW detector network configurations and minimum frequencies in LIGO Livingston. The differences between runs and their corresponding figures are presented in Tab.~\ref{tab:Bilbyrunsettings}.

%%%%%%%%%%%%%%%%%%%%%%%%%%%%%%%%%%%%%%%%%%%%%%%
\subsection{{\tt BayesWave} CBC and glitch analyses}
\label{appendix:bayeswave}
%%%%%%%%%%%%%%%%%%%%%%%%%%%%%%%%%%%%%%%%%%%%%%%

%
\begin{table}[]
\begin{tabular}{c @{\quad\quad} c @{\quad\quad}c }
 Figure(s) & Models &  Detector Network  \\
 \hline
 \ref{fig:BWBilbyinV},~\ref{fig:SNRV} & CBC+glitch & HLV \\
 \ref{fig:BWBilbyinL},~\ref{fig:GlitchDraws} & CBC+glitch & HL \\
 \ref{fig:SNRV} & glitch & V \\
 \ref{fig:LLOGlitchCompare} & glitch & L \\
\hline
\hline
\end{tabular}
\caption{Table of {\tt BayesWave} runs and settings. All analyses use 4\,s of data, a low frequency cut-off of $\flow=20$\,Hz, a sampling rate of 2048\,Hz, and the {\tt IMRPhenomD} waveform when the CBC model is used. Furthermore, all analyses use the original strain data without the glitch mitigation described in Sec.~\ref{appendix:glitchsubtraction}. Columns correspond to the main text figures each analysis appears in, the {\tt BayesWave} models that are used, and the detector network (H: LIGO Hanford, L: LIGO Livingston, V: Virgo). While not plotted in any figure, we also performed ``CBC+Glitch" analyses on injections into the HL detector network as a glitch background study on GW200129-like sources, see Sec.~\ref{sec:glitchL}.}
\label{tab:BWrunsettings}
\end{table}

{\tt BayesWave}~\cite{Cornish:2014kda,Littenberg:2014oda,Cornish:2020dwh} is a flexible data analysis algorithm that models combinations of coherent generic signals, glitches, Gaussian noise, and most recently, CBC signals that appear in the data~\cite{Hourihane:2022doe, Chatziioannou:2021ezd, Wijngaarden:2022sah}. To sample from the multi-dimensional posterior for all the different models, {\tt BayesWave} uses a ``Gibbs sampler" which cycles between sampling different models while holding the parameters of the non-sampling model(s) fixed. 

For this analysis, we mainly use the CBC and glitch models (a setting we refer to as ``CBC+Glitch"). The CBC model parameters (see App.~\ref{appendix:bilby}) are sampled via a fixed-dimension Markov Chain Monte Carlo sampler (MCMC) using the priors described in Wijngaarden \textit{et al.}~\cite{Wijngaarden:2022sah}. The glitch model is based on sine-Gaussian wavelets and samples over both the parameters of each wavelet (central time, central frequency, quality factor, amplitude, phase~\cite{Cornish:2014kda}) and the number of wavelets via a trans-dimensional or Reverse-jump MCMC. In some cases, we also make use of solely the glitch model (termed ``glitchOnly" analyses) that assumes no CBC signal and the excess power is described only with wavelets.
The differences between runs and the figures in which they appear are presented in Tab.~\ref{tab:BWrunsettings}.

Though {\tt BayesWave} typically marginalizes over uncertainty in the noise PSD~\cite{Littenberg:2014oda}, in this work we use the same fixed PSD as the {\tt Bilby} runs for more direct comparisons. Additionally, we use identical data as App.~\ref{appendix:bilby} for the LIGO Hanford and Virgo detectors. However, when it comes to LIGO Livingston we use the original (i.e., ``unmitigated," without any glitch subtraction) data in order to independently infer the glitch. We do not marginalize over uncertainty in the detector calibration.

%%%%%%%%%%%%%%%%%%%%%%%%%%%%%%%%%%%%%%%%%%%%%%%
\section{Select results with {\tt IMRPhenomXPHM}}
\label{app:systematics}
%%%%%%%%%%%%%%%%%%%%%%%%%%%%%%%%%%%%%%%%%%%%%%%

\begin{figure}
    \centering
    \includegraphics[width=\linewidth]{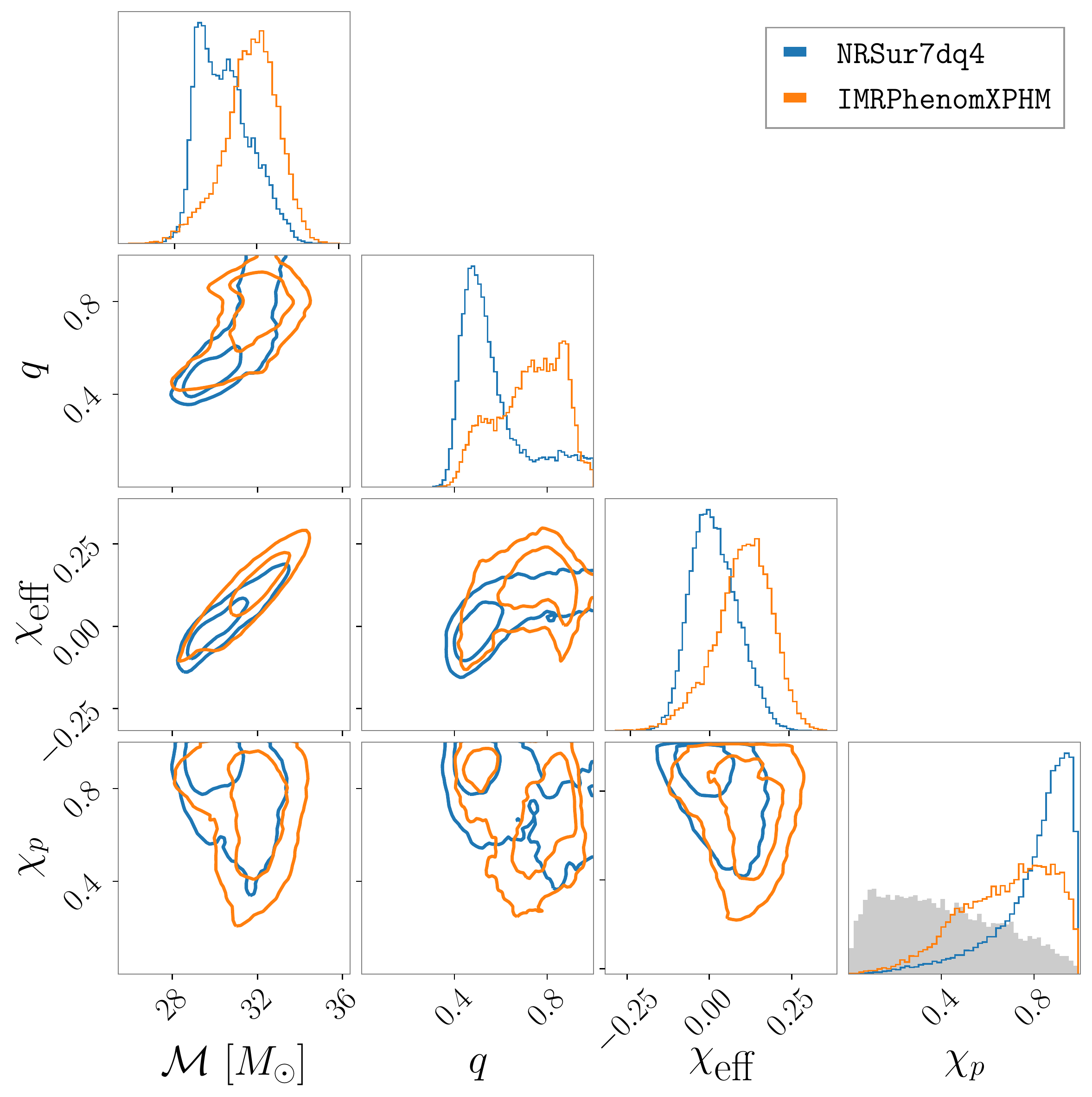}
    \caption{Similar to Fig.~\ref{fig:intrinsic}, using data from LIGO Livingston and LIGO Hanford. The comparison shows slight tension between results when using {\tt NRSur7dq4} and {\tt IMRPhenomXPHM}, though qualitatively {\tt IMRPhenomXPHM} also seems to support the evidence for spin-precession.}
    \label{fig:waveforms}
\end{figure}

In this Appendix, we present select results obtained with the {\tt IMRPhenomXPHM}~\cite{Pratten:2020ceb} waveform model that also resulted in evidence for spin-precession in GWTC-3~\cite{gwtc3}. Even though {\tt IMRPhenomXPHM} and {\tt NRSur7dq4} both support spin-precesion, in contrast to {\tt SEOBNRv4PHM}, there are still noticeable systematic differences between them. Figure~\ref{fig:waveforms} shows that while {\tt NRSur7dq4} and {\tt IMRPhenomXPHM} generally have overlapping regions of posterior support, {\tt IMRPhenomXPHM} shows slightly more preference for higher $q$ and less support for extreme precession when compared to {\tt NRSur7dq4}. Waveform systematics are expected to play a significant role in GW200129's inference (e.g. Refs.~\cite{gwtc3,Hannam:2021pit, Hu:2022rjq}), which motivates utilizing {\tt NRSur7dq4} for all of our main text results.

\bibliography{OurRefs}

\end{document}